\documentclass[twocolumn]{aastex63}
\pdfoutput=1 
\usepackage{amsmath,amstext}
\usepackage[T1]{fontenc}
\usepackage{apjfonts} 
\usepackage[figure,figure*]{hypcap}
\usepackage{longtable}

\shorttitle{Elusive AGNs with JWST }
\shortauthors{S. Satyapal et al.}

\begin{document}

\title{The Diagnostic Potential of JWST in Characterizing Elusive AGNs}

\author[0000-0003-2277-2354]{Shobita Satyapal}
\affiliation{George Mason University, Department of Physics and Astronomy, MS3F3, 4400 University Drive, Fairfax, VA 22030, USA}
\author{Lara Kamal}
\affiliation{George Mason University, Department of Physics and Astronomy, MS3F3, 4400 University Drive, Fairfax, VA 22030, USA}
\author[0000-0003-1051-6564]{Jenna M. Cann}
\affiliation{George Mason University, Department of Physics and Astronomy, MS3F3, 4400 University Drive, Fairfax, VA 22030, USA}
\affiliation{National Science Foundation, Graduate Research Fellow}

\author[0000-0002-4902-8077]{Nathan J. Secrest}
\affiliation{U.S. Naval Observatory, 3450 Massachusetts Avenue NW, Washington, DC 20392-5420, USA}

\author{Nicholas P. Abel}
\affiliation{MCGP Department, University of Cincinnati, Clermont College, Batavia, OH 45103, USA}

\begin{abstract}

It is now clear that a significant population of accreting black holes are undetected by commonly employed optical, mid-infrared color, X-ray, and radio selection methods due to obscuration or contamination of the emission from the nuclear region caused by star formation in the host galaxy. These elusive active galactic nuclei (AGNs) are an important component of the AGN population. They are often found in mergers, where significant black hole growth likely takes place, and in the lowest mass galaxies or galaxies that lack classical bulges, a
demographic that places important constraints on models of supermassive black hole seed formation and merger-free models of AGN fueling. In this work, we demonstrate the power of {\it James Webb Space Telescope (JWST)} in uncovering and characterizing these elusive AGNs. We present an integrated modeling approach in which both the line and emergent continuum is predicted from gas exposed to the ionizing radiation from a young starburst and an AGN, focusing specifically on the spectral diagnostics available through {\it JWST}, and provide predictions on the detectability of key diagnostic lines by the NIRSpec and MIRI spectrometers, assuming typical conditions for the gas. We demonstrate the crucial need for {\it JWST} in uncovering low level accretion activity even in nearby galaxies and out to redshifts of $\approx$ 1 - 3, depending on the ionization parameter, and gas phase metallicity. We present a redshift-dependent selection strategy that can be used to identify promising elusive AGN candidates for future follow-up study. The full suite of simulations are available online, where users can select specific parameters and retrieve the predicted continuum and infrared line luminosities from our models.  

\end{abstract}


\section{Introduction}

Detecting complete samples of active galactic nuclei (AGNs) and understanding their connection to the properties of the host galaxies in which they reside has been an important goal of extragalactic astronomy. This is because we have come to realize that supermassive black holes are a fundamental component of galaxies, and that they play a vital role in their evolution. When accreting, AGNs can exercise a profound effect over the baryonic component of the universe by driving massive energetic outflows that can redistribute dust and metals into the intergalactic medium \citep[e.g.,][]{veilleux2005,veilleux2020}, quench star formation in the host galaxy \citep[e.g.,][]{choi2015,somerville2008}, and impact the size and structure of the host galaxies \citep[e.g.,][]{magorrian1998, gebhardt2000, gultekin2009, mcconnell2013}. 
\par

Unfortunately, detecting complete samples of AGNs and characterizing their properties has been challenging since accretion signatures are often missed by optical spectroscopic, broad-band mid-infrared, X-ray, and radio surveys due either to obscuration of the nuclear region, or contamination of the emission from star formation in the host galaxy \citep[e.g.,][]{condon1991,goulding2009,delmoro2013,trump2015, satyapal2018}. These widely used surveys are even more limited in low metallicity galaxies and in galaxies with lower mass black holes, where contamination by star formation is even more significant at optical and X-ray wavelengths \citep{trump2015,fragos2013}, and accretion signatures are missed by optical narrow emission line ratios \citep{groves2006,cann2019}. Even optical broad lines, usually considered a hallmark signature of an AGN, are often found to be associated with supernova activity in low mass galaxies \citep{baldassare2016}. Finally, surveys based on variability studies are even more limited in the hunt for elusive AGNs and depend strongly on the cadence and baseline of the observations (Secrest \& Satyapal 2020).
\par
The limitations of these previous widely used diagnostics in finding and characterizing elusive AGNs is a significant obstacle. This is  because these elusive AGNs are often found in key phases of galaxy evolution, such as late stage galaxy mergers, when the black hole is expected to grow most rapidly \citep{vanwassenhove2012,blecha2013} but be accompanied by enhanced star formation and obscuration \citep{satyapal2017, blecha2018,pfeifle2019}, or in low mass and bulgeless galaxies, a galaxy population that may place important constraints on models of SMBH seed formation and merger-free models of SMBH growth \citep[e.g.,][]{volonteri2009, volonteri2010, vanwassenhove2010, greene2012}. Moreover, only $\approx$ 1\% of galaxies are actively accreting. Studying accretion properties in the low luminosity regime is essential in order to gain insight into the connection between active and quiescent galaxies, and the phase in galaxy evolution when the AGN ''turns off''.
\par

Observations of the infrared coronal lines (CLs) are a promising tool to identify elusive AGNs and characterize their properites.  This is because even hot massive stars do not produce photons with energies sufficient to produce the ions, and infrared CL emission from Type II supernovae is rare, extremely weak, and short lived \citep[e.g.,][]{benjamin1990,greenhouse1990,smith2009}, demonstrating that the detection of a prominent line confirms the presence of an AGN. Thus apart from being less sensitive to obscuration than optical emission lines, infrared CLs are insensitive to contamination from star formation in the host galaxy, and the detection of a single line with a given luminosity alone is sufficient to confirm the presence of an AGN. The power of these diagnostics in finding elusive AGNs has been strikingly demonstrated by the {\it Infrared Space Observatory (ISO)} and the {\it Spitzer} missions through the discovery of a population of AGNs in galaxies that display optically "normal" nuclear spectra \citep[e.g.,][]{lutz1999,genzel1998, satyapal2007,satyapal2008,satyapal2009,goulding2009, veilleux2009}. 
\par
Apart from their diagnostic potential in finding elusive AGNs, infrared coronal lines can be used to indirectly reconstruct the intrinsic spectral energy distribution (SED) from the Lyman limit up to several hundred electron volts, a region of the electromagnetic spectrum that is observationally inaccessible because of Galactic and intrinsic absorption \citep[e.g.,][]{moorwood1996,alexander2000}. The extinction-insensitive emission lines arise from collisionally ionized forbidden fine structure transitions from highly ionized species, with ionization potentials that extend well beyond the Lyman limit to several hundred electron volts, corresponding to the so-called ``Big Blue Bump'' thought to arise from the accretion disk emission \citep[e.g.,][]{lyndenbell1969,shakura1973,netzer1985} and the ``soft excess'' characteristic of AGN soft X-ray spectra \citep{krolik1999}. The line ratios can therefore in principle be used to constrain this part of the electromagnetic spectrum and discriminate between various accretion models, and even possibly the mass of the black hole \citep{cann2018}. The resulting intrinsic SED is also of interest in studying the effect of the AGN on its environment, a crucial input in understanding AGN feedback and its impact on star formation in the host galaxy. Not only do infrared fine structure lines carry the advantage of being less sensitive to dust extinction, the excitation energies corresponding to the transitions are small compared to the ambient nebular temperature, and so the infrared line ratios of two different stages of ionization produced by the same element are only weakly dependent on the electron temperature of the gas in which they are produced. The line ratios are also an ideal tool to be used to constrain the gas density and, together with the optical emission lines, the extinction and gas temperature of the optically obscured and highly ionized gas around the AGN.
\par
With the advent of the {\it James Webb Space Telescope (JWST)}, infrared spectroscopic observations with unprecedented sensitivity in the $1-30~\mu$m  wavelength range will be possible, enabling the observation of emission lines that were never detected by {\it Spitzer} and {\it ISO}. Moreover, high resolution spectroscopic capabilities were only available beyond $10~\mu$m with {\it Spitzer}.  Even in the brightest nearby galaxies, only a limited number of emission lines in the $1-30~\mu$m  wavelength range have been detected, and the full diagnostic potential of emission lines in the $1-30~\mu$m wavelength range has not been explored. The goal of this paper is to explore the diagnostic potential of these emission lines in the study of elusive AGNs and to provide predictions for the estimated brightness of these lines for a range of models. Toward that end, we conducted a theoretical investigation of the emission line spectrum and the mid-infrared SED produced by gas ionized and dust heated by an AGN and a starburst stellar population. These models employ an integrated modeling approach using photoionization and stellar population synthesis models  in which both the line and emergent continuum is predicted from gas and dust exposed to the ionizing radiation from a young starburst and an AGN. In a previous paper \citep{satyapal2018}, we presented the mid-infrared SED predicted by our models. 
Here we present the emergent infrared emission line spectrum from our models and assess their diagnostic potential in the study of elusive AGNs in {\it JWST} era. We make our model grids publicly available online for use by the community.

\par

This paper is organized as follows. In Section 2, we describe our model, stating all of our assumptions on the input radiation field, the metallicity, geometry, total column density, and gas properties. In Section 3, we discuss our main results, focusing on the line luminosities predicted in the {\it JWST} range as a function of our model parameters, and present the predicted brightest expected emission lines that will be detected. In Section 4, we provide a redshift-depended pre-selection strategy that can be employed to identify promising AGN candidates for follow-up study. In Section 5, we summarize our results.

\section{Theoretical Calculations}

We performed a series of calculations using the spectral synthesis code \textsc{Cloudy} version c17 \citep{ferland2017}.  The ability of \textsc{Cloudy} to both model the emission-line spectrum along with a detailed model of dust \citep{vanHoof2004} makes it an ideal computational tool to study the sensitivity of the gas and dust spectrum to the input ionizing radiation field and the physical conditions of the gas.  In addition, \textsc{Cloudy} has the ability to compute the spectrum of multiple gas phases, including H II regions, photodissociation regions (PDRs), and molecular clouds, in a single calculation, by assuming an equation of state between each gas phase (usually constant density or constant pressure).  A complete list of physical processes treated by \textsc{Cloudy} in each physical regime can be found in \citet{ferland2013}. The shape of the emergent infrared dust continuum and the emission line spectrum from gas irradiated by a given radiation field depends on several factors, including the shape of the ionizing radiation field, the geometrical distribution of the gas and the stopping criterion, the chemical composition and grain properties, the gas properties and the equation of state. The model assumptions we employ are discussed in depth in \citep{satyapal2018}. Here we give a summary and highlight all major assumptions. The full set of simulated SEDs and emission line luminosities for the brightest lines in the $1-30~\mu$m  wavelength range are available to download online  \footnote{\url{https://bgc.physics.gmu.edu/jwst-spectral-predictor/}}. We note that we have not included the effects of shocks in our calculations.  CLs can also be produced in hot ( $\approx 10^{6}$ K) collisionally ionized plasma \citep[e.g.,][]{oke1968,nussbaumer1970}, however several studies suggest that the main driver of the CL emission is photoionization by high energy radiation \citep[e.g.,][]{oliva1994,whittle2005,schlesinger2009,mazzalay2010} with possibly some contribution from shocks \citep[e.g.,][]{tadhunter1988,morse1996,rodriguez2006, geballe2009}. The contribution to the CL region from shocks is beyond the scope of this work. Note that the strength and shape of the ionizing radiation field produced by shocks is a strong function of the shock velocity  with CL line emission generated only for a narrow range of parameter space with luminosities that are much lower than can be produced through direct photoionization by the ionizing source \citep{contini1997}. Indeed observations of nearby Herbig-Haro objects and supernova remnants (SNRs), where there are no strong, external sources of ionizing radiation, suggest that while high ionization emission lines indicate the presence of ionizing shocks, the UV continuum flux is two orders of magnitude lower than that produced from a power law ionizing radiation field \citep{morse1996}. In addition, a significant contribution from shocks would be accompanied by high shock velocities which would result in other distinguishing features in the emission line spectrum \citep{allen2008,kewley2013}. 

\subsection{Incident Radiation Field}
 
As in \citet{satyapal2018}, our calculations are designed to model the emergent line and continuum emission from gas illuminated by  a mix of a starburst and AGN SED.  We have adopted a more complex model for the AGN than employed in \citet{satyapal2018} since the detailed shape of the high energy radiation field from the AGN has a critical impact on the CL emission line spectrum. We adopt the approach we employed in \cite{cann2018}. In brief, for the AGN, we assume that the AGN continuum consists of three components: an accretion disk, Comptonized X-ray radiation in the form of a power law, and an additional component seen in the X-ray spectrum of most AGNs, often referred to as the ``soft excess component". Multiwavelength observations of quasars suggest that the continuum emission from AGNs peaks in the ultraviolet part of the electromagnetic spectrum \citep[e.g.,][]{shields1978,elvis1986,laor1990}. This emission, often referred to as the "Big Blue Bump"(BBB) is attributed to the emission from the accretion disk around the black hole. We adopt a simple geometrically thin, optically thick accretion disk model \citep{shakura1973}, where the emission from the disk is approximated by the superposition of blackbodies at temperatures corresponding to the  different disk annuli at radius R, with the temperature as a function of radius R given by \citet{peterson07, frank2002}:

\begin{equation} \label{Eq 1}
T = 6.3 \times 10^{5} \bigg( \frac{\dot m}{\dot m_{Edd}} \bigg)^{1/4} \bigg( \frac{M_{BH}}{10^{8}M_{\odot}} \bigg)^{-1/4}\bigg( \frac{R}{R_s} \bigg)^{-3/4} K
\end{equation}

where $\dot m_{Edd}$ and $\dot m$ are the Eddington and actual accretion rates, respectively, M$_{BH}$ is the mass of the black hole, and R$_s$ is the Schwarzchild radius. 
In this study, the accretion disk spectrum was modeled using the {\it diskbb} model \citep{mitsuda1984, makishima1986} of XSpec v12.9.0\footnote{\url{http://www.heasarc.gsfc.nasa.gov/docs/xanadu/xpec}}  (\cite{arnaud96}). We adopt a black hole mass of 10$^{7.5}$ M$_{\odot}$ and assumed that the black hole is accreting at 0.1~$\dot m_{Edd}$. Note that varying the black hole mass or accretion rate will change the shape of the AGN SED, which we do not explore in this paper. Given the adopted SED, we fix the total luminosity of our model. The effects of varying the black hole mass on the infrared emission line spectrum are explored in \citet{cann2018}. We also include a high energy power law component to the AGN SED, believed to be caused by Comptonization of seed photons produced by the disk \citep{svensson1999} by energetic electrons in a hot corona \citep{krolik1999}.  We assume a spectral index of 0.8 \citep[e.g.,][]{wilkes1987, grupe2006} and an $\alpha_{OX}$ ratio of 1.2 \citep{netzer1993} We note that since the emission lines we are analyzing arise from ions with ionization potentials of less than or equal to 400 eV, the detailed shape of the AGN SED at the highest energies will not impact our calculations. We assume a high energy cutoff of 100~keV. Finally,  we include in our  models a ''soft excess'' component, modeled phenomenologically as a blackbody with a temperature of 150~eV.  Note that because the ionization potentials of the coronal lines extend into the soft X-rays, it is important to include this component into our AGN continuum. The details of our approach can be found in \citet{cann2018}. 

For the starburst input SED, we used a 5 Myr continuous star-formation starburst model  from the Starburst99 website\footnote{\url{http://www.stsci.edu/science/starburst99/docs/parameters.html}} \citep{leitherer99,Leitherer14} as our input SED with a Saltpeter initial mass function (IMF) of power-law index $\alpha=-2.35$ and a star-formation rate of 1 ~$M_\sun$~yr$^{-1}$, with lower and upper mass cutoffs of 1 and 100 $M_\sun$, respectively. All other parameters were left at their default settings from the Starburst99 website. These settings produce the hardest radiation field from a stellar population as discussed in \citet{satyapal2018}. Since our goal here is to determine the emission lines that can be used to confirm the presence of an AGN and characterize its properties, our choice of starburst SED was governed by requiring the most extreme ionizing radiation field from a stellar population.  In that way, we can asses which emission lines can robustly be assumed to require the presence of an AGN. As discussed in \citet{satyapal2018}, the emergence of Wolf-Rayet stars at 5 Myr, along with massive O stars results in the hardest radiation field from a starburst \citep{schaerer1996,kehrig2008}. This extreme starburst model employed in this work will produce emission lines from ions with the highest ionization potentials possible from purely stellar processes.

The total luminosity is fixed at $L=2\times10^{44}$~erg~s$^{-1}$.  Note that the total luminosity and star formation rate do not affect the shape of the ionizing radiation field for our assumed  extreme starburst model, and line luminosities can multiplied by an appropriate scale factor. We allow the fraction of the total luminosity that is due to AGN and starburst activity to vary, such that the total luminosity is fixed but the relative contribution of each SED varies.  Cosmic rays and the Cosmic Microwave Background (CMB) are both included, with the cosmic ray ionization rate set to $3~\times10^{-16}$~ s$^{-1}$ \citep{indriolo2012}. As in \citet{satyapal2018}, we performed all calculations using two separate metallicities. The first set of models assume solar metallicity, both for the stellar population as well as the interstellar medium (ISM).  For the lower metallicity models we use the SED for the 0.10 models from Starburst99, keeping the age the same as before, and reducing the gas metallicity by the same amount.  Note that we consistently vary both the metallicity of the stellar population as well as the metallicity of the gas in our models. For the AGN models, only the ISM abundance is changed between the high and low metallicity models, where the abundance set and depletion factors are matched with the starburst models. It is possible that the varying the metallicity may affect the torus and accretion disk structure surrounding the black hole, altering the radiation field incident on the narrow line region. However, since the effect of metallicity on the intrinsic AGN SED is unknown, we assume the same incident ionizing radiation field for all AGN models presented in this work.

\begin{figure}
\includegraphics[width=0.45\textwidth]{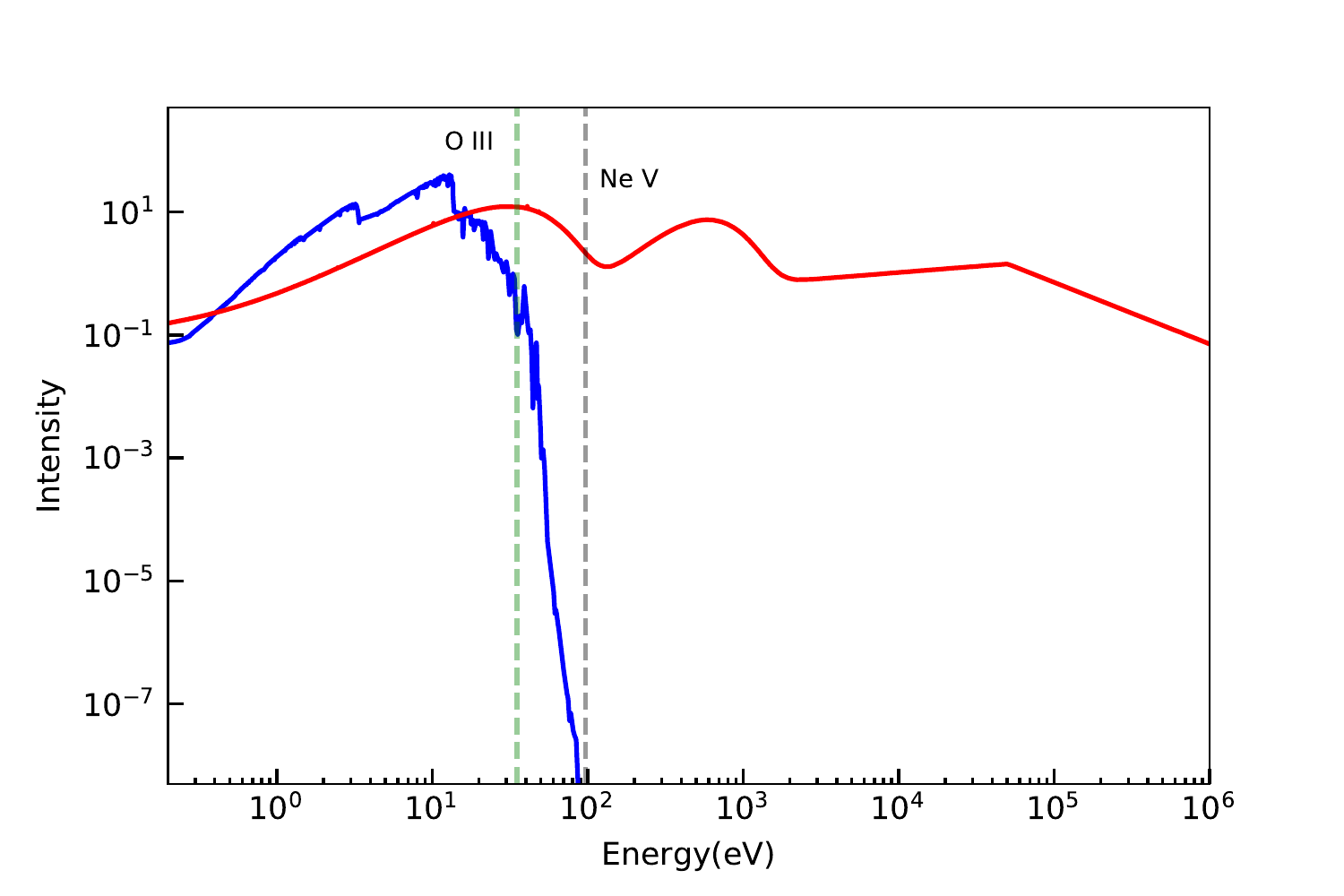}
\includegraphics[width=0.45\textwidth]{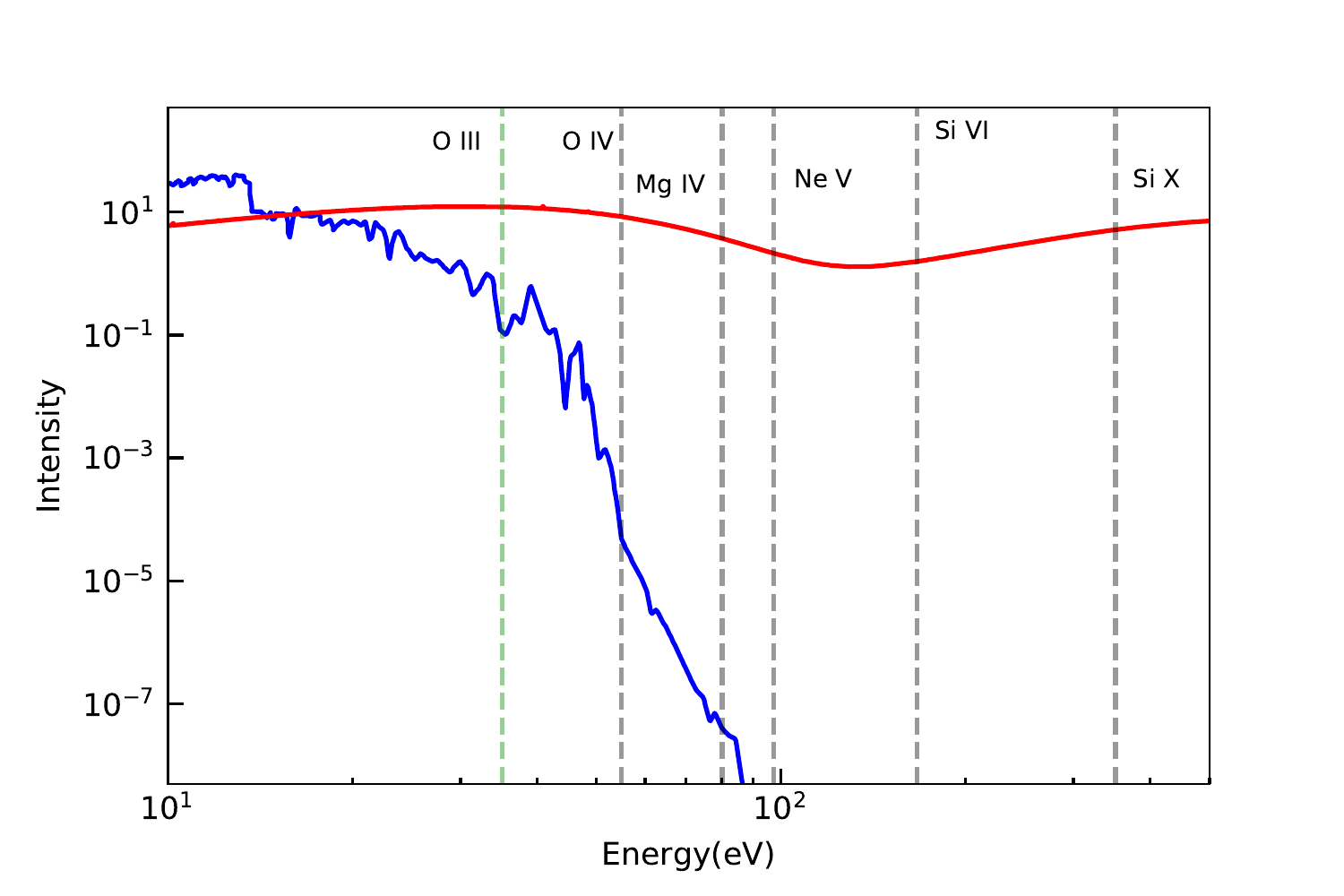}
\caption{Top Panel: Incident  SED for the solar metallicity starburst model \textbf{(blue line)} and AGN SED \textbf{(red line)}. The dotted vertical lines indicate the ionization potentials of O~III and Ne~V ions. Bottom Panel: Same as above, showing only the UV to soft X-ray component of the SEDs. The dotted vertical lines indicate the ionization potentials corresponding to a selection of coronal lines observable in the infrared. The corresponding ion is labeled in the figure. }
\label{SED}
\end{figure}

In Figure ~\ref{SED}, we plot the incident continuum for our solar metallicity extreme starburst and AGN model.  As can be seen, the incident continuum of the starburst drops dramatically in the UV part of the electromagnetic spectrum, producing an insignificant fraction of the total luminosity beyond a few tens of eV. In contrast, the SED of the AGN, which includes the soft excess and power law components, extends well past the SED of the starburst into the high energy X-ray part of the electromagnetic spectrum. In the top panel of Figure ~\ref{SED}, we also display the ionization potentials of O~III and Ne~V. As can be seen, photoionization from the ionizing radiation field for a starburst will be a strong contributor to the optical [O~III] emission line luminosities, limiting the diagnostic potential of this line in finding elusive AGNs due to significant contamination by star formation. However, as can be seen, photoionization by an extreme starburst cannot contribute to the production of any ions with ionization potentials above $\approx$ 100 ~eV. Thus infrared emission lines such as the [Ne~V] $14~\mu$m line is a robust indicator of an AGN, even in the presence of an extreme starburst and significant extinction. In the bottom panel of Figure ~\ref{SED}, we show the ionization potential of various other ions with emission lines in the infrared part of the electromagnetic spectrum. As can be seen, these lines sample the AGN SED  well into the soft X-rays and can be used to gain insight into the otherwise unobservable intrinsic radiation field produced by the AGN.

\subsection{Geometrical Distribution of the Gas and Microphysical Assumptions}

We assume a one-dimensional spherical model with a closed geometry, where the cloud is between the observer and the continuum source, and the ionization parameter and gas density are allowed to vary.  The ionization parameter, $U$, is defined as the dimensionless ratio of the incident ionizing photon density to the hydrogen density:
\begin{equation}
U=\frac{\phi_H}{n_H}=\frac{Q(H)}{4\pi R^{2}n_{H}c}
\end{equation}

where $\phi_H$ is the flux of hydrogen-ionizing photons striking the illuminated face of the cloud per second, $n_{H}$ is the Hydrogen density, $c$ is the speed of light, and $Q(H)$ is the number of hydrogen ionizing photons striking the illuminated face per second. In our calculations, $U$ is allowed to vary from  $10^{-4}$ to $10^{-1}$.   On average, $U$ is $\sim10^{-3}$ based on observations of the optical BPT emission lines in local galaxies and HII regions \citep{dopita2000, moustakas2010} or quasars \citep{baskin2005}, but can be higher in ULIRGS \citep{abel09} or high-redshift galaxies \citep{pettini01, brinchmann08, maiolino2008, hainline2009, erb2010}, and in the CL region \citep{ferguson1997,alexander1999}. We therefore explored a large dynamic range of ionization parameters in this work.
\par

We chose the gas and dust abundances in our calculations to be consistent with the local interstellar medium (ISM). We point the reader to \citet{satyapal2018} for a full description of our dust model. Note that for the low metallicity models, we scale all metals along with the dust abundance by a factor of 0.10. We assume constant density, and vary the Hydrogen density of the gas from $\log(n_\textrm{H}/$cm$^{3})$ = $2.5-3$, in units of 0.5 dex. We include thermal, radiation, and turbulent pressure in our models with the turbulent velocity set to 5.0 km s$^{-1}$.   The assumption of constant pressure means the density will increase with depth into the cloud \citep{pellegrini2007} as the calculation reaches the hydrogen ionization front and moves into the PDR.  Each calculation is computed for two iterations, which is required for optically thick models in order to compute the emergent spectrum. 
\par
Our calculations end for a range of total hydrogen column densities $N(H)$, so we can study the effects of extinction on the emergent emission line strengths and the shape of the continuum. The stopping criterion (corresponding to a particular $A_V$) more significantly impacts the optical emission line strengths. Apart from the attenuation of the emission line luminosities, higher column densities result in
 cooler dust for a given ionizing radiation field, which in turn affects the shape of the mid-infrared continuum, the depth of the 9.7~\micron\ silicate feature, and the strength of the PAH features since the abundance of PAHs varies with cloud depth as we discuss in greater detail in \citet{satyapal2018}. We allow $N(H)$ to vary from $10^{19}$ cm$^{-2}$ to $10^{24}$ cm$^{-2}$, in increments of 0.5 dex.  The upper limit of the column density considered by our calculations assure that, for a given U, the calculations extend beyond the hydrogen ionization front and into the PDR/molecular cloud.
\par
Given the full suite of composite SEDs, $n_\textrm{H}$, $N(H)$, radius, metallicity, and equation of state, we computed a total of 3,328 simulations. For each model, we calculated the emergent emission line spectrum in the $1-30~\mu$m  range and list the brightest expected emission lines from the ionized gas for our range of models in Table 1. Note that in this paper, we focus exclusively on the emission lines from the ionized gas, and not those from the atomic, molecular, and photodisocciated gas. The line luminosities for all lines for our full range of parameter space, including those from the PDR, are available online, as are the full transmitted SEDs.

\startlongtable
\begin{deluxetable*}{lcccc}
\tabletypesize{\footnotesize}

\tablecaption{Brightest Emission Lines from the Ionized Gas in the $1-30~\mu$m Range}

\tablehead{
\colhead{Line} & \colhead{Transition} & \colhead{Wavelength} & \colhead{Ionization Potential} & \colhead{Critical Density}\\
 \colhead{}& \colhead{}&   \colhead{($\mu$m)} & \colhead{(eV)} & \colhead{(cm$^{-3}$)}
 }
 \decimals
\startdata
\lbrack Fe XIII\rbrack* &  $^3$P$_0 - ^3$P$_1$ & 1.0747 & 331 & $6.498 \times 10^8$ \\ 
 \lbrack S IX\rbrack* &  $^3$P$_2 - ^3$P$_1$ & 1.2520 &  328 & $2.68\times10^9$ \\ 
  \lbrack Si X\rbrack* &  $^2$P$_{1/2} - ^2$P$_{3/2}$ & 1.4301 & 351 & $1.300 \times 10^8$ \\ 
 Pa $\alpha$* & $4-3$  & 1.87510  & 14 & $...$
 \\
  \lbrack Si XI\rbrack* &  $^3$P$_1 - ^3$P$_2$ & 1.93446 & 523 & $1.129 \times 10^8$ 
  \\ 
 Br $\delta$*  & $8-4$ & 1.94454 & 14 & $...$
 \\ 
 \lbrack Si VI\rbrack* &  $^2$P$_{1/2} - ^2$P$_{3/2}$ & 1.963410 & 167 & $3.022 \times 10^8$ \\ 
 \lbrack Al IX\rbrack* & $^2$P$_{1/2} - ^2$P$_{3/2}$ & 2.0450 & 285 & $9.806 \times 10^7$
 \\ 
 Br $\gamma$* &  $7-4$ & 2.16551  & 14 & $...$ \\ 
 \lbrack Ca VIII\rbrack* & $^2$P$_{1/2} - ^2$P$_{3/2}$ & 2.32117 & 127 & $5.012 \times 10^6$ \\ 
 \lbrack Si IX\rbrack* &  $^3$P$_{2} - ^3$P$_{1}$ & 2.580 & 304 & $4.668 \times 10^7$ \\ 
 \lbrack Fe IV\rbrack &  $^4$D$_{1/2} - ^4$P$_{3/2}$ & 2.80554 & 31 & $2.942 \times 10^6$ \\ 
 \lbrack Fe IV\rbrack & $^4$D$_{7/2} - ^4$P$_{5/2}$  & 2.83562 & 31 & $2.235 \times 10^3$ \\ 
 \lbrack Fe IV\rbrack & $^4$D$_{1/2} - ^4$P$_{1/2}$  & 2.86447  & 31 & $7.268 \times 10^6$ \\ 
 \lbrack Mg VIII\rbrack* & $^2$P$_{1/2} - ^2$P$_{3/2}$  & 3.027950  & 225 & $1.61\times 10^7$ \\ 
 D I & $14-6$ & 4.01974 & 14 & $...$ \\ 
 Br $\alpha$* & $5-4$ & 4.05113 & 14 & $...$ \\   
\lbrack Mg IV\rbrack* & $^2$P$_{1/2} - ^2$P$_{3/2}$ & 4.48712 & 80 & $1.285 \times 10^7$ \\ 
\lbrack Ar VI\rbrack* & $^2$P$_{1/2} - ^2$P$_{3/2}$ & 4.52800 & 75 & $7.621 \times 10^5$  \\   
\lbrack K III\rbrack & $^2$P$_{1/2} - ^2$P$_{3/2}$ & 4.61683 &  32 & $1.238 \times 10^6$ \\ 
\lbrack Fe II\rbrack* & $^4$F$_{9/2} - ^6$D$_{9/2}$ & 5.33881 & 8 & $1.078 \times 10^8$ \\ 
\lbrack Mg VII\rbrack & $^3$P$_2-^3$P$_1$ & 5.51 & 186 & $4.580\times 10^6$ \\ 
 \lbrack Mg V\rbrack* & $^3$P$_{2} - ^3$P$_{1}$  & 5.60700 & 109 & $5.741 \times 10^6$ \\ 
 \lbrack Ar III\rbrack &$^3$P$_{0} - ^3$P$_{2}$ & 6.36676 & 28 & $4.698 \times 10^1$ \\ 
 \lbrack Ar II\rbrack* & $^2$P$_{1/2} - ^2$P$_{3/2}$ & 6.98337 & 16 & $4.192 \times 10^5$ \\ 
 \lbrack Na III\rbrack* & $^2$P$_{1/2} - ^2$P$_{3/2}$ & 7.31706 & 47 & $3.039 \times 10^6$ \\ 
 \lbrack Ne VI\rbrack* & $^2$P$_{3/2} - ^2$P$_{1/2}$ & 7.64318 & 126 & $3.573 \times 10^5$ \\ 
 \lbrack Ar V\rbrack & $^3$P$_{1} - ^3$P$_{2}$ & 7.89971 & 60 & $2.025 \times 10^5$ \\ 
 \lbrack Ar III\rbrack* & $^3$P$_{1} - ^3$P$_{2}$ & 8.98898 & 28 & $3.491 \times 10^5$ \\ 
 \lbrack S IV\rbrack* & $^2$P$_{3/2} - ^2$P$_{1/2}$ & 10.5076 & 35 & $4.254 \times 10^4$ \\ 
 \lbrack Cl IV\rbrack & $^3$P$_2 - ^3$P$_1$ & 11.7629 & 40 & $7.665 \times 10^4$ \\ 
 \lbrack S III\rbrack &  $^3$P$_2 - ^3$P$_0$ & 12.0004 & 23 & $1.281 \times 10^0$ \\ 
 \lbrack Ne II\rbrack* & $^2$P$_{1/2} - ^2$P$_{3/2}$ & 12.8101 & 22 & $6.341 \times 10^5$ \\ 
 \lbrack Ar V\rbrack & $^3$P$_{0} - ^3$P$_{1}$ & 13.0985 & 60 & $9.516 \times 10^4$ \\ 
 \lbrack Ne V\rbrack* & $^3$P$_{1} - ^3$P$_{2}$ &  14.3228 & 97 & $4.792 \times 10^4$ \\ 
 \lbrack Cl II\rbrack* & $^3$P$_{1} - ^3$P$_{2}$ &  14.367 & 13 & $5.531 \times 10^4$ \\ 
 \lbrack Ne III\rbrack* & $^3$P$_{1} - ^3$P$_{2}$ &  15.5509 & 41 & $3.388 \times 10^5$ \\ 
 \lbrack S III\rbrack* & $^3$P$_{2} - ^3$P$_{1}$ & 18.7078 & 23 & $1.609 \times 10^4$\\ 
 \lbrack Fe VI\rbrack & $^4$F$_{3/2} - ^4$F$_{5/2}$ & 19.5527 & 75 & $1.456 \times 10^5$ \\ 
 \lbrack Cl IV\rbrack & $^3$P$_{1} - ^3$P$_{0}$ & 20.3197 & 40 & $4.035 \times 10^4$ \\ 
 \lbrack Ar III\rbrack* & $^3$P$_{0} - ^3$P$_{1}$ & 21.8253 & 28 & $4.770 \times 10^4$ \\ 
 \lbrack Fe III\rbrack* & $^5$D$_3-^5$D$_4$ & 22.925 & 16 & $9.876 \times 10^4$ \\ 
 \lbrack Ne V\rbrack* & $^3$P$_{0} - ^3$P$_{1}$ & 24.2065 & 97 & $2.656 \times 10^4$ \\ 
 \lbrack O IV\rbrack* & $^2$P$_{3/2} - ^2$P$_{1/2}$ & 25.8832 & 55 & $1.429 \times 10^4$ \\ 
 \lbrack Fe V\rbrack & $^5$D$_{3} - ^5$D$_{2}$ & 25.9131 & 55 & $1.095 \times 10^5$ \\
\enddata
\label{tab1}
\end{deluxetable*}

\section{Predicted Infrared Emission Line Spectrum}

\begin{figure*}[tb]
\centering
\includegraphics[width=0.8\textwidth]{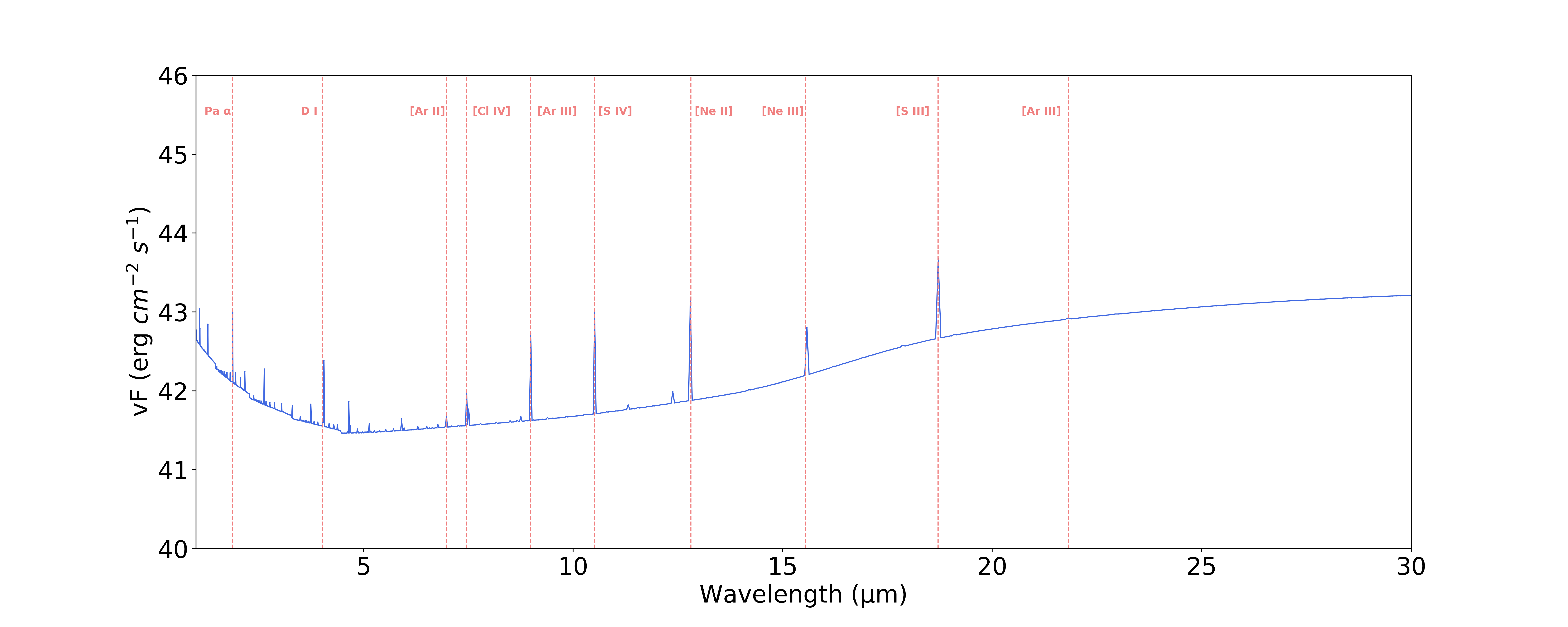}
\includegraphics[width=0.8\textwidth]{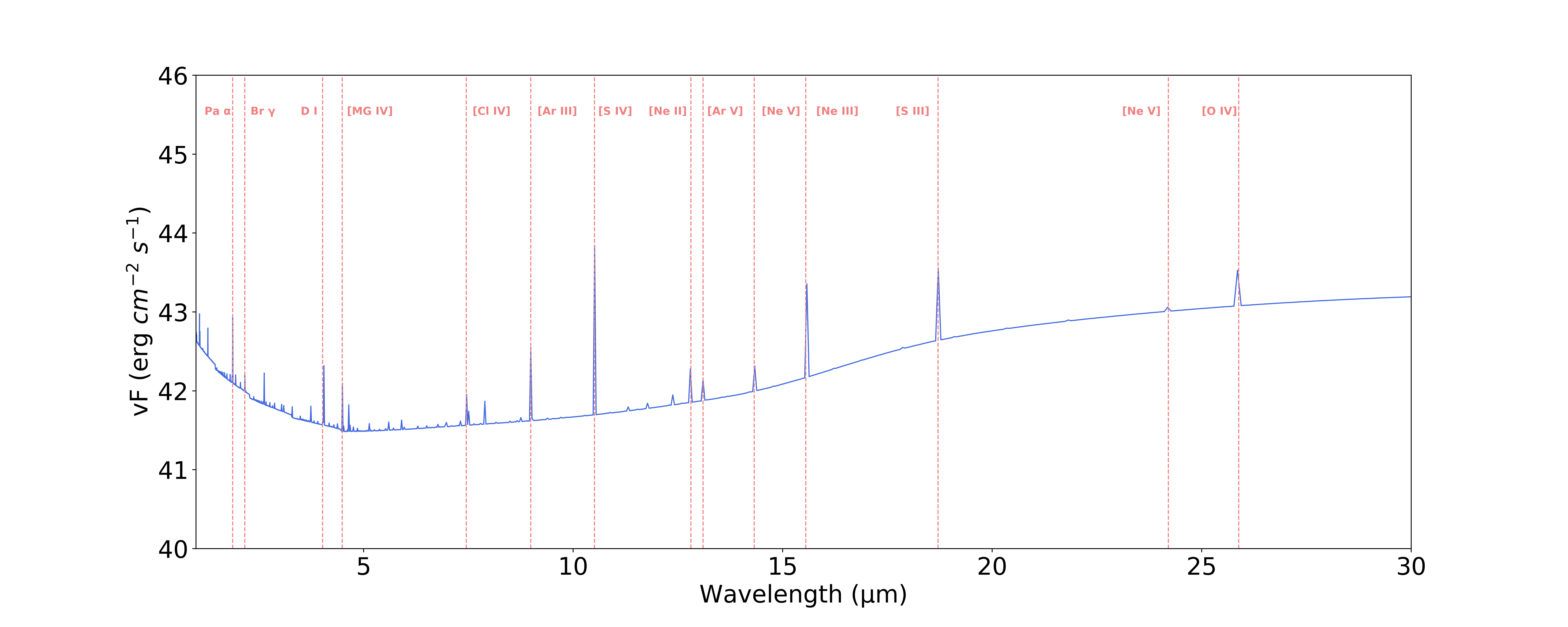}
\includegraphics[width=0.8\textwidth]{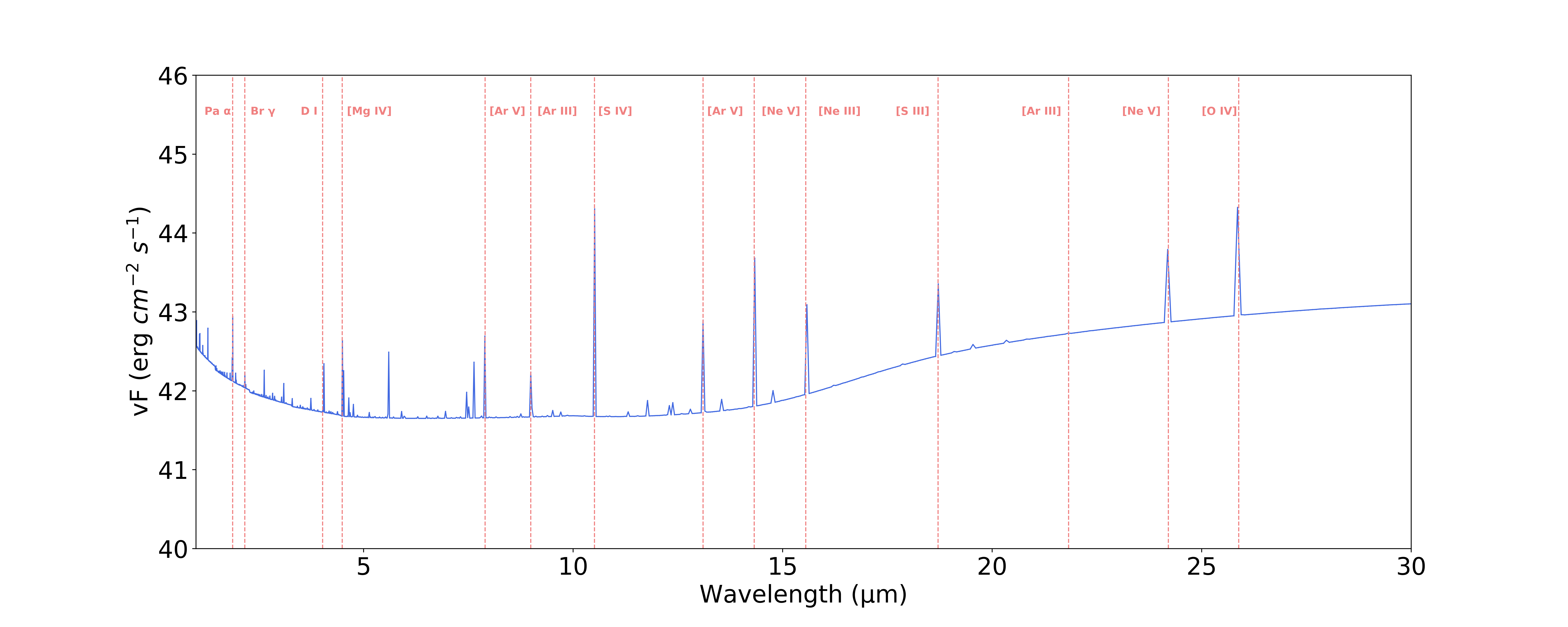}
\includegraphics[width=0.8\textwidth]{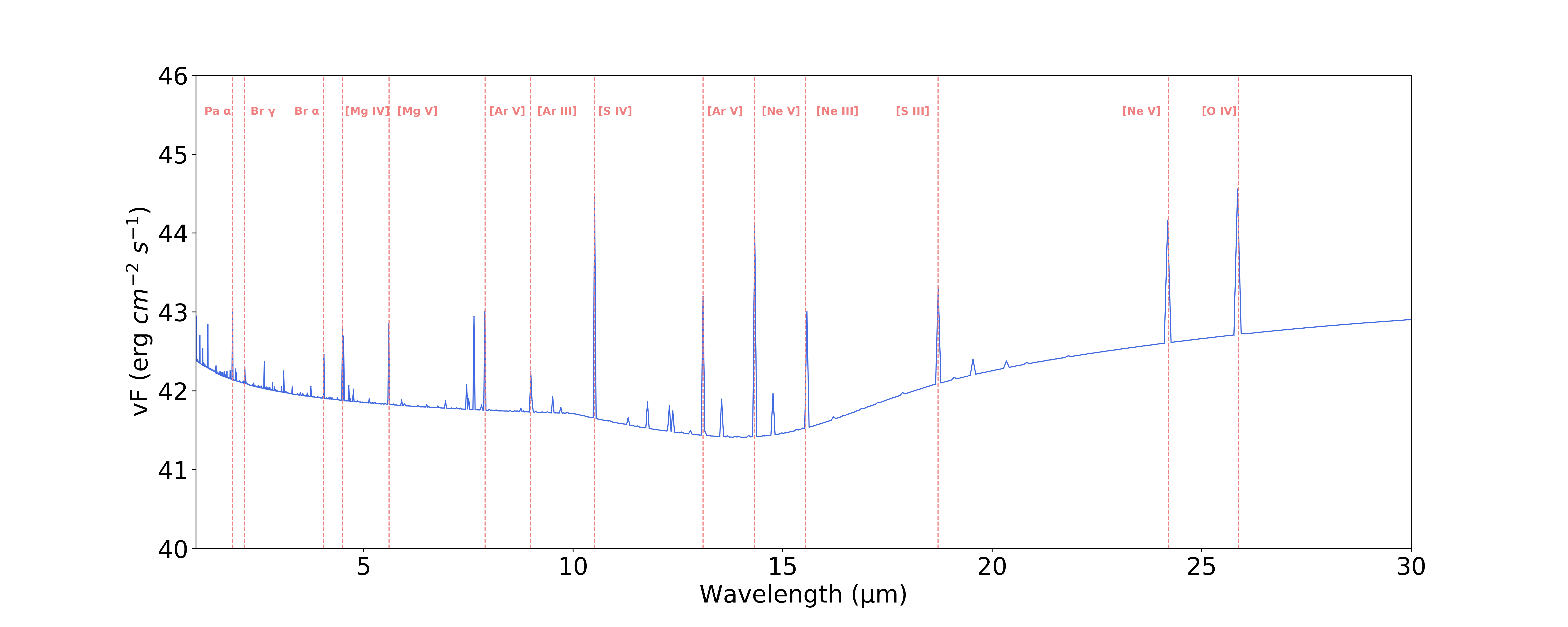}
\caption{Simulated spectra between $1-30$ $\micron$ for a few representative composite SEDs \textbf{(from top to bottom: 0\%, 5\%, 40\%, 100\%)} for our solar metallicity models with a standard gas density of $n_{\textrm{H}}=300~\rm{cm^{-3}}$, column density of $\log(N(H)/\textrm{cm}^{-2})=21$, and an ionization parameter of $\log U=-2.0$. Key emission lines are labeled in the figure.}
\label{spectra}
\end{figure*}

\begin{figure}
\includegraphics[width=0.45\textwidth]{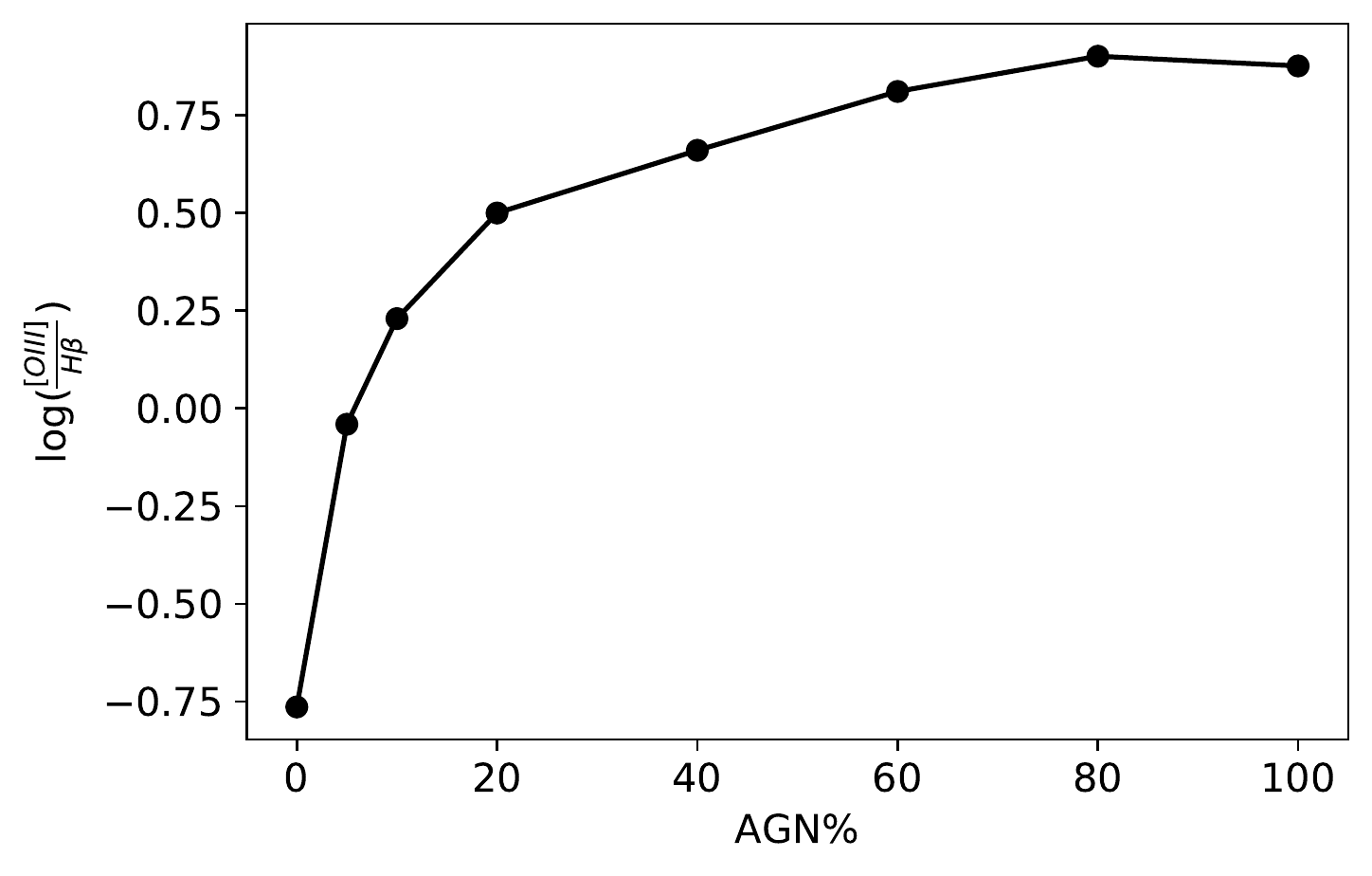}
\caption{[O~III]/H$\beta$ luminosity ratio as a function of \% AGN for our solar metallicity model with a standard gas density of $n_{\textrm{H}}=300~\rm{cm^{-3}}$, column density of $\log(N(H)/\textrm{cm}^{-2})=21$, and an ionization parameter of $\log U=-3.0$.  }
\label{optical_line_lum}
\end{figure}

\begin{figure}
\includegraphics[width=0.45\textwidth]{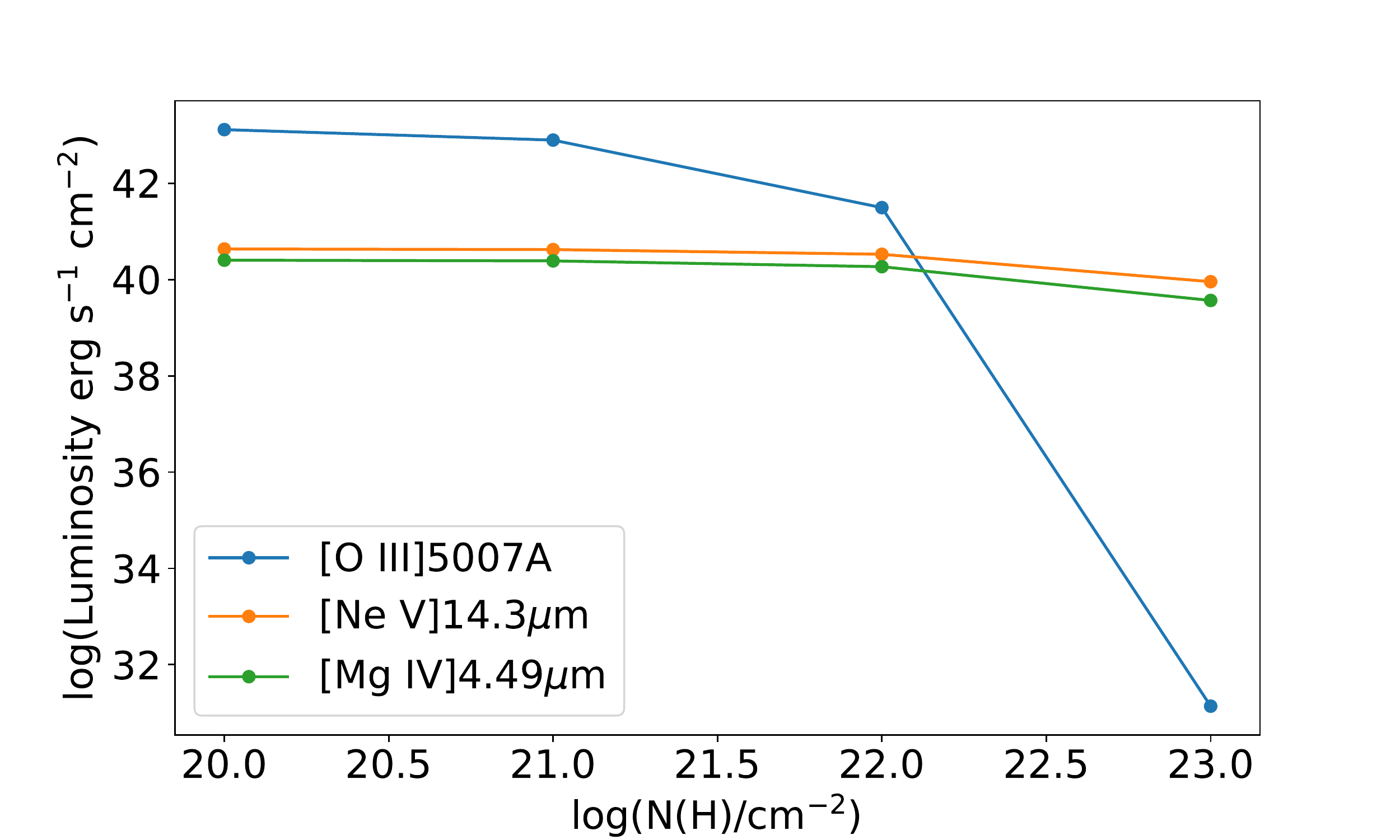}
\caption{A selection of line luminosities a function of column density for our solar metallicity 100\% AGN model with a standard gas density of $n_{\textrm{H}}=300~\rm{cm^{-3}}$. Note that plotted line luminosities correspond to a total bolometric lumimosity of  $L=2\times10^{44}$~erg~s$^{-1}$, and an ionization parameter of $\log U=-3.0$.  }
\label{NH}
\end{figure}

The predicted emergent infrared emission line spectrum in our simulations for our  $\log U=-2.0$, $n_{\textrm{H}}=300~\textrm{cm}^{-3}$, $\log(N(H)/\textrm{cm}^{-2})=21$ solar metallicity model for a few representative composite SEDs, in order of increasing AGN contribution, are shown in Figure ~\ref{spectra}. As can be seen, emission lines from ions with high ionization potentials are missing in the pure starburst model but become prominent as the AGN radiation field dominates the ionizing SED with the relative strengths of the various emission lines changing along the sequence displayed in Figure ~\ref{spectra}. While emission lines from ions with lower ionization potentials from, for example, Ne~II, Ne~III, S~III, S~IV, Ar~II, Ar~III and the hydrogen recombination lines,  are ubiquitous and prominent in the full range of models, the emission lines from ions with higher ionization potential, such as Mg~IV, Mg~V, Ne~V, Ne~VI, etc. are produced only when an AGN radiation field is included. As can be seen from Figure \ref{SED}, at the ionization potential of O~III (35~eV), there is a substantial contribution from photoionization from stars making it impossible to identify low level accretion activity. Figure \ref{optical_line_lum} shows the variation of the [O~III]/H$\beta$ luminosity ratio, used in the most widely used optical AGN diagnostic (commonly referred to as the Baldwin-Phillips-Terlevich, BPT, diagram) as a function of increasing AGN fraction for our $\log U=-3.0$, $n_{\textrm{H}}=300~\textrm{cm}^{-3}$, $\log(N(H)/\textrm{cm}^{-2})=21$ solar metallicity model.  Note that these are typical parameters based on observations of the optical emission lines in nearby galaxies and HII regions \citep{dopita2000, moustakas2010}, but the line luminosities for the full grid can be obtained online. As can be seen for low values of \% AGN, the [O~III]/H$\beta$ ratio does not meet the AGN classification criteria for widely used optical narrow line diagnostics \citep{kewley2001}, and therefore it cannot identify AGNs dominated by starbursts in contrast to the infrared emission lines even in the absence of obscuration. The detection of a single line with high ionization potential alone can confirm the presence of the AGN even when it contributes only 5\% of the bolometric luminosity, highlighting the power of infrared spectroscopy in identifying AGNs and characterizing their properties in a regime where optical diagnostics fail. Furthermore, unlike the optical emission lines, the infrared lines are insensitive to extinction even for extreme column densities. Figure \ref{NH} shows the optical [OIII] 5007 emission line luminosity as a function of the column density for our 100\% AGN solar metalicity model using a standard gas density and ionization parameter. As can be seen, the optical emission line luminosity drops to undetectable levels for column densities in excess of $\log(N(H)/\textrm{cm}^{-2})=22$, while the infrared coronal line luminosities remain largely unchanged up to column densities of $\log(N(H)/\textrm{cm}^{-2})=23$, underscoring the power in infrared diagnostics in uncovering and characterizing obscured AGNs.

\par
\begin{figure}
\includegraphics[width=0.45\textwidth]{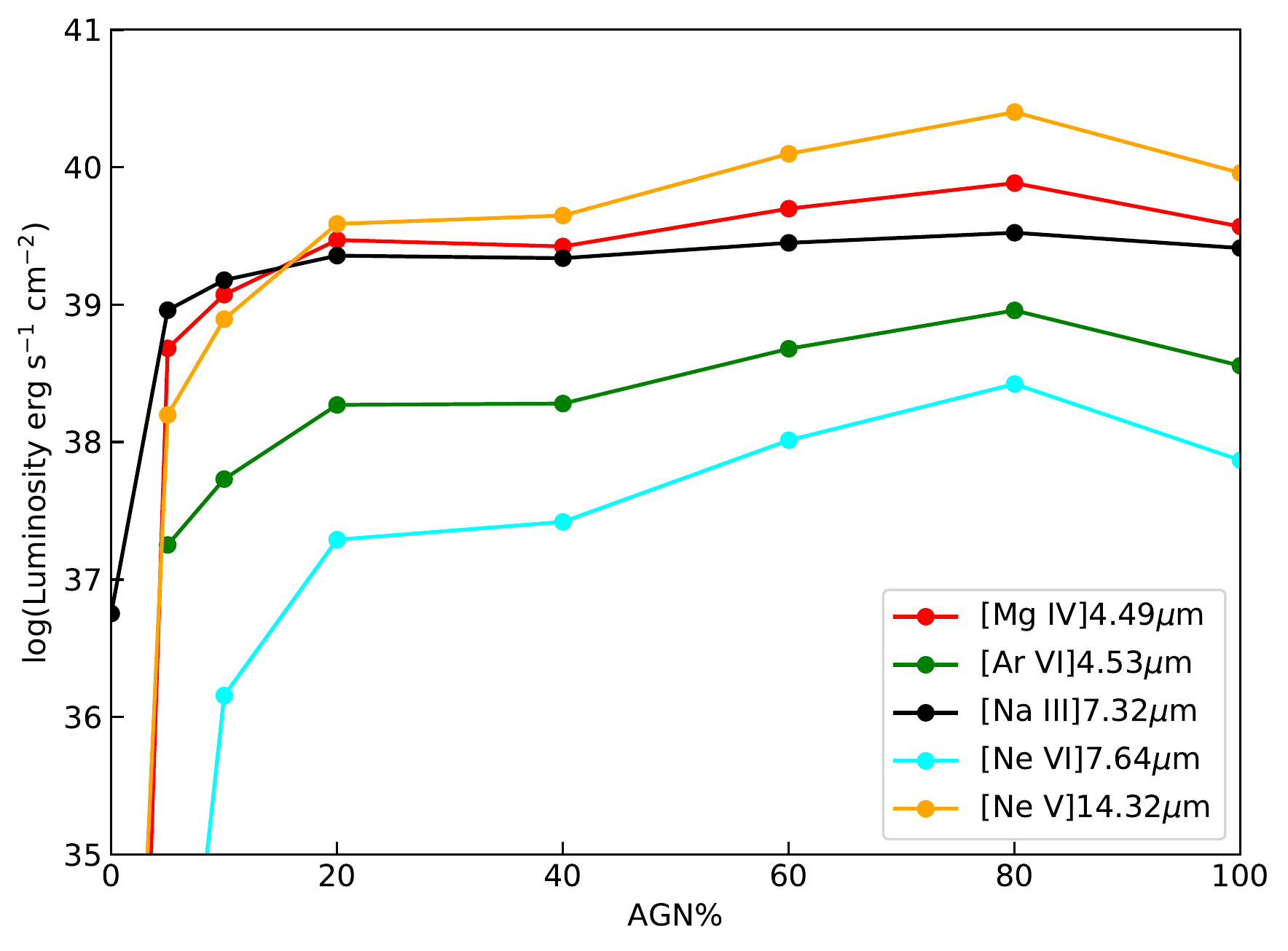}
\includegraphics[width=0.45\textwidth]{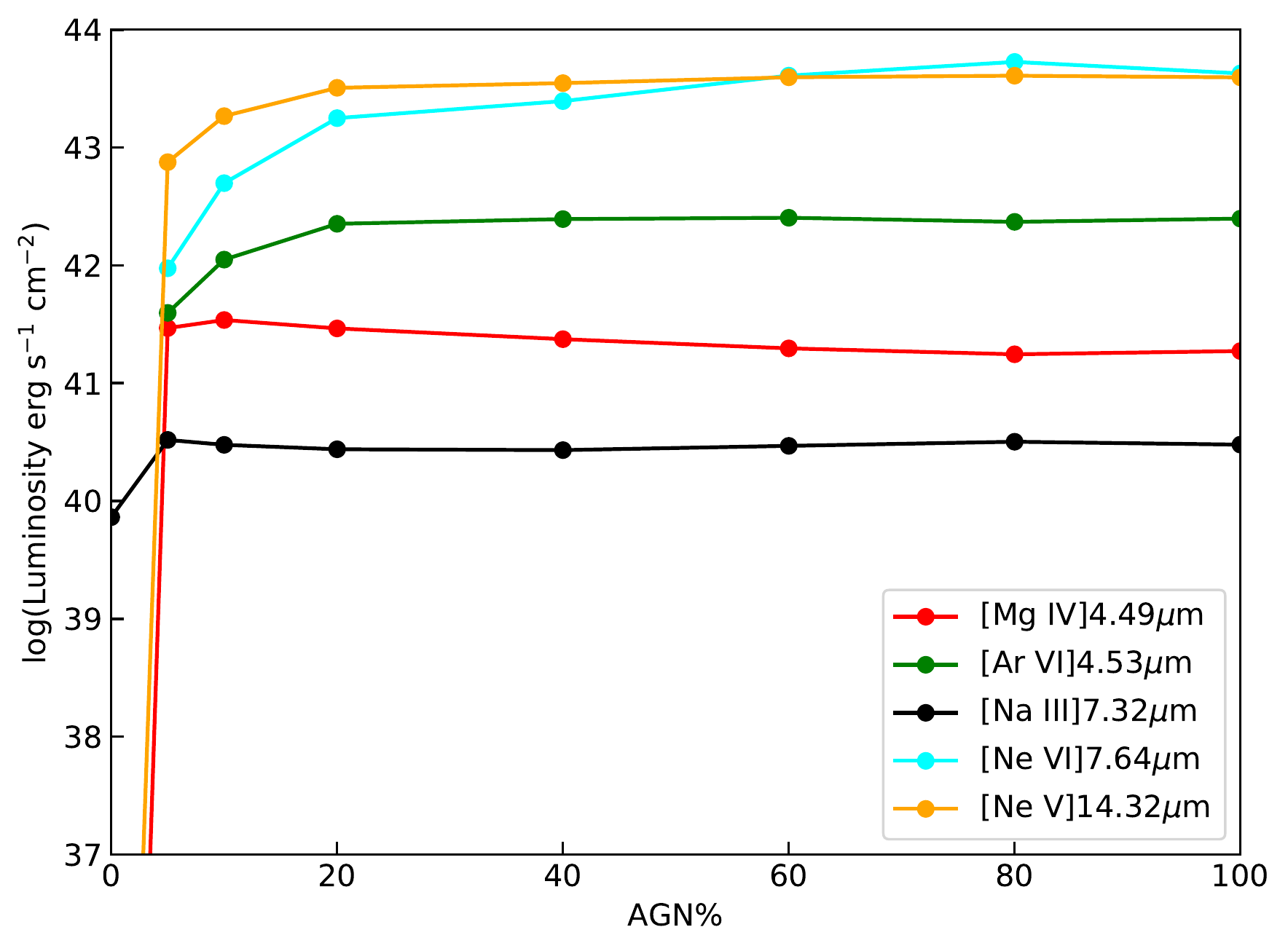}
\caption{Top Panel: A selection of bright infrared emission lines in the MIRI wavelength range as a function of \% AGN for for our solar metallicity mdoel with a standard gas density of $n_{\textrm{H}}=300~\rm{cm^{-3}}$, column density of  $\log(N(H)/\textrm{cm}^{-2})=23$, total bolometric lumimosity of  $L=2\times10^{44}$~erg~s$^{-1}$, and an ionization parameter of $\log U=-3.0$. Bottom Panel: Same as above but showing our $\log U=-1.0$ model, with all other parameters unchanged. }
\label{line_lum}
\end{figure}
\begin{figure}

\includegraphics[width=0.45\textwidth]{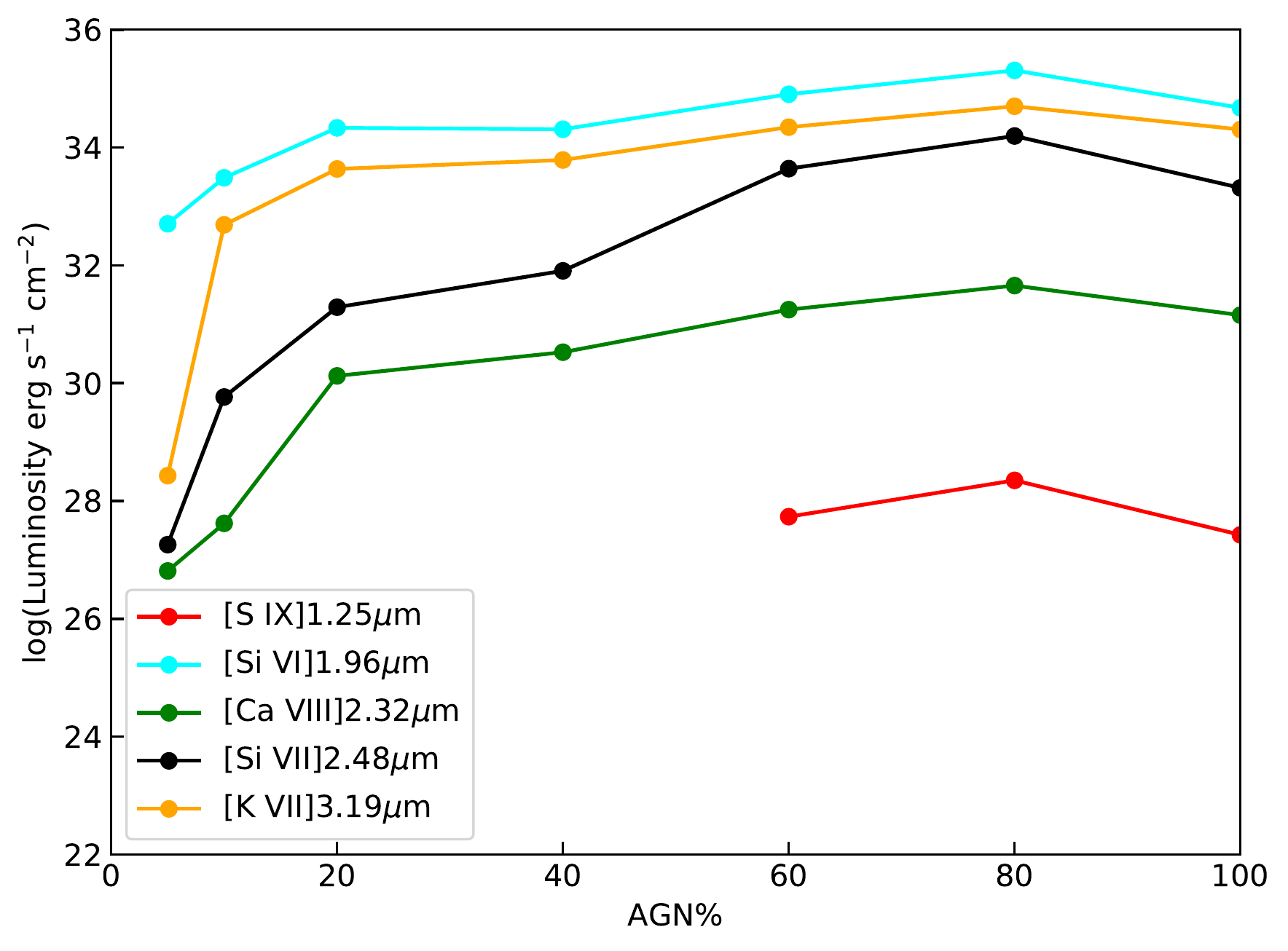}
\includegraphics[width=0.45\textwidth]{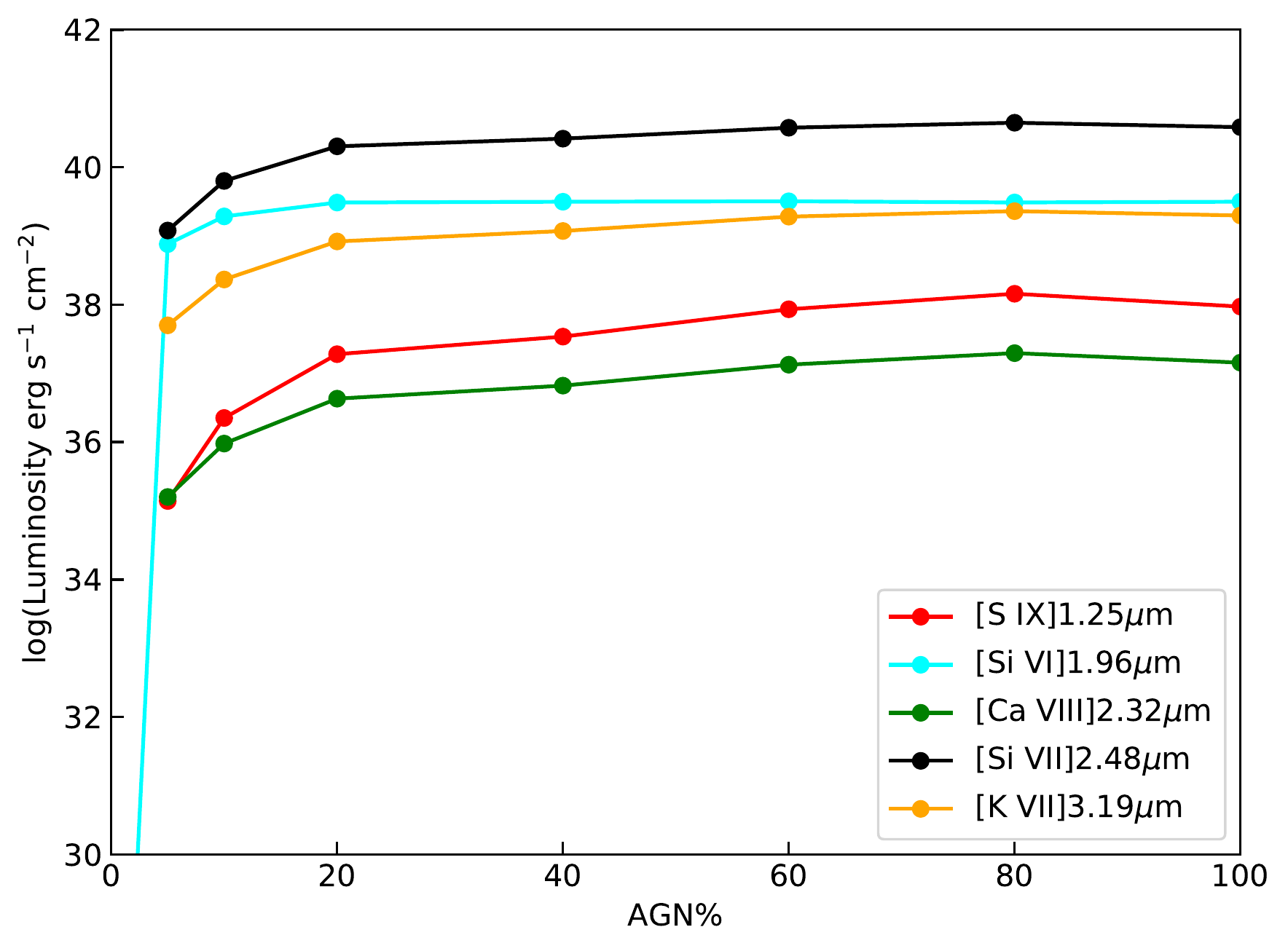}
\caption{Top Panel: A selection of bright emission lines in the NIRSpec range as a function of \% AGN for our solar metallicity model with a standard gas density of $n_{H}=300~\rm{cm^{-3}}$, column density of \textbf{$\log(N(H)/\textrm{cm}^{-2})$ =23}, total bolometric lumimosity of  $L=2\times10^{44}$~erg~s$^{-1}$, and an ionization parameter of $\log U=-3.0$. Bottom Panel: Same as above but showing our $\log U=-1.0$ model, with all other parameters unchanged.Note that some of the lines are truncated since for the highest ionization potential lines, a significant AGN contribution is required to produce the emission line.}
\label{line_lum_nirspec}
\end{figure}

Figure \ref{line_lum} shows various prominent infrared emission lines accessible to the MIRI instrument on {\it JWST} as a function of increasing AGN fraction for  the $n_{\textrm{H}}=300~\textrm{cm}^{-3}$, $\log(N(H)/\textrm{cm}^{-2})=23$ solar metallicity model for two different ionization parameters.  From Figure \ref{line_lum}, we see that for low level accretion activity, i.e., when the starburst dominates the bolometric luminosity, the brightest lines, assuming the gas is dominated by lower ionization parameters, will be the [Mg~IV]4.49~$\mu$m and [Na~III]7.32~$\mu$m emission lines, with the [Ne~V]14.3~$\mu$m becoming the most prominent line at higher levels of accretion activity. For gas that is dominated by higher ionization parameters, the [Ne~V]14.3~$\mu$m line will be prominent at all levels of activity, with the [Ne~VI]7.64~$\mu$m line becoming prominent only when the AGN dominates the bolometric luminosity of the galaxy. Figure \ref{line_lum_nirspec} shows several prominent emission lines in the wavelength range of NIRSpec for the same model parameters displayed in Figure \ref{line_lum}. In this wavelength range, the most luminous diagnostic line is typically the [Si~VI]1.96~$\mu$m line, detected in nearby bright AGNs from the ground \citep{lamperti2017, rodriguez2006}. The models predict that the diagnostic lines in the NIRSpec range that identify elusive AGNs are intrinsically several magnitudes fainter than those in the MIRI range. In addition, at the highest column densities, unlike the mid-infrared lines, the near-infrared lines are attenuated at the highest column densities, highlighting the importance of the mid-infrared coronal lines in finding and characterizing the most obscured accretion activity.

\section{Detectability of Key Lines by {\it JWST}}

\begin{figure}

\includegraphics[width=0.45\textwidth]{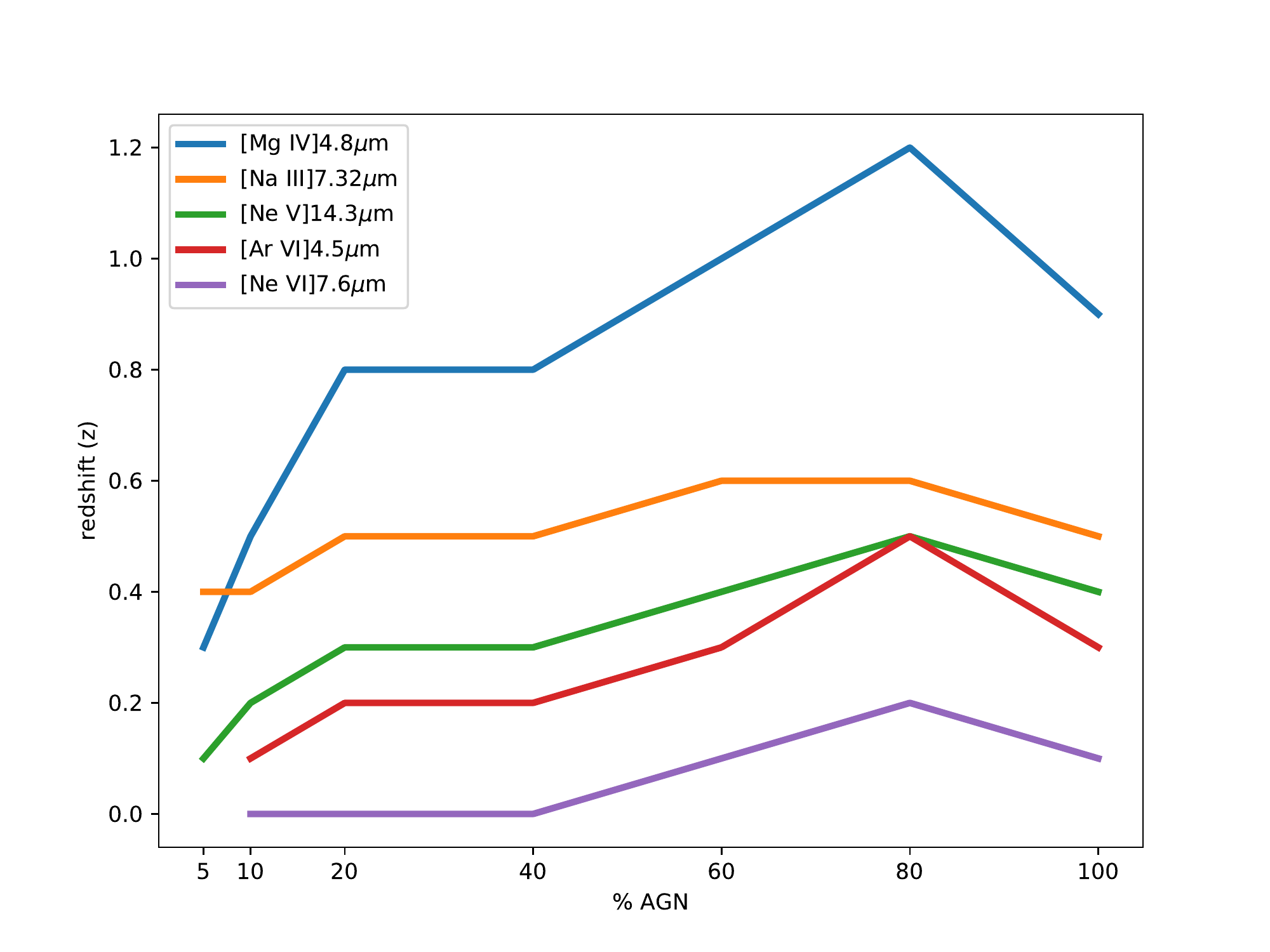}
\includegraphics[width=0.45\textwidth]{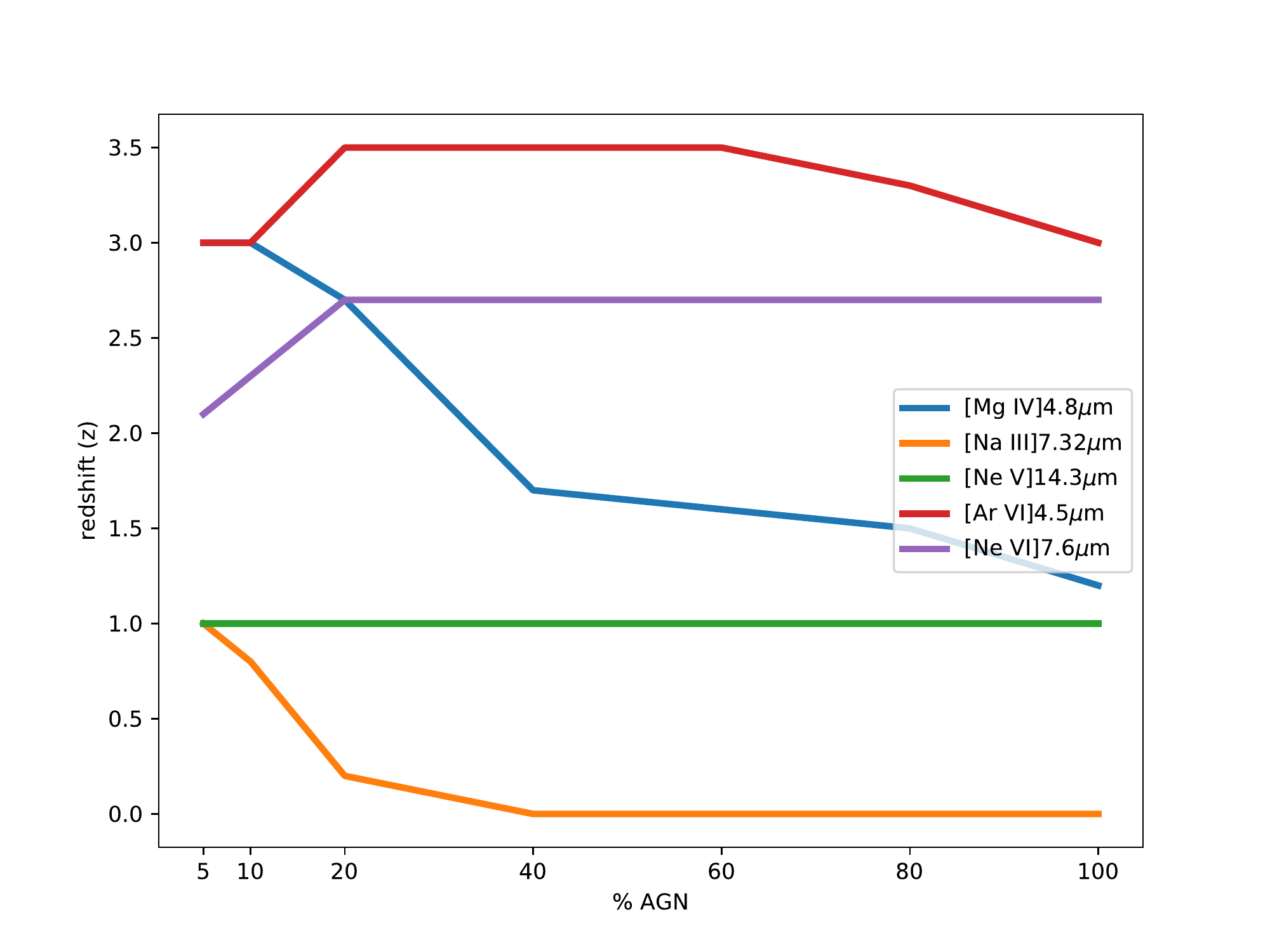}
\caption{Top Panel: Redshift at which key diagnostic lines with rest wavelengths in the MIRI wavelength range can be detected assuming predicted sensitivities as a function of \% AGN for our solar metallicity model with a standard gas density of $n_{\textrm{H}}=300~\rm{cm^{-3}}$, \textbf{column density of $\log(N(H)/\textrm{cm}^{-2})=23$}, total bolometric lumimosity of  $L=2\times10^{44}$~erg~s$^{-1}$}, and an ionization parameter of $\log U=-3.0$. Bottom Panel: Same as above but showing our $\log U=-1.0$ model, with all other parameters unchanged.
\label{lum_miri}
\end{figure}

We explored the detectability of the various key diagnostic lines by {\it JWST} based on our model predictions using the predicted sensitivity limits of MIRI and NIRSpec\footnote{MIRI sensitivity limits are taken from \citep{glasse2015}, and NIRSpec and Gemini R=1000 curves taken from \url{http://www.stsci.edu/jwst/about-jwst/history/historical-sensitivity-estimates}. Gemini sensitivities are assumed for NIRI with R=1000. Detection thresholds correspond to a 10$\sigma$ sensitivity in a 10,000 second exposure}. These calculations are based on our assumed total luminosity of $L=2\times10^{44}$~erg~s$^{-1}$, which is comparable to nearest Seyfert galaxy, Circinus \citep{maiolino1998}. Figure \ref{lum_miri}, shows the redshift at which the lines with rest wavelength in the MIRI wavelength range can be detected as a function of \%AGN. For low ionization parameters, low level AGN activity is detectable with these diagnostic lines only out to a redshift of approximately 0.4, while dominant AGNs are detectable by these lines up to a redshift of 1.2. Note that while the [Ne~V]14.3~$\mu$m line is more luminous than many of the diagnostic emission lines as shown in Figure \ref{line_lum}, the sensitivity of the instrument is worse at these wavelengths, making the [Mg~IV]4.49~$\mu$m the line of choice. At high ionization parameters, the [Mg~IV] and [Ar~VI] lines can be detected out to a redshift 3.0 for low level activity, and up to a redshift of 3.5 for dominant AGNs. Note that for the high ionization parameter models, while the [Ne~VI] and [Ne~V] lines are more luminous, they are redshifted out of the MIRI band at redshifts of $\approx$2.7 and 1, respectively, limiting the use of these emission lines for higher redshift targets.

\begin{figure}

\includegraphics[width=0.45\textwidth]{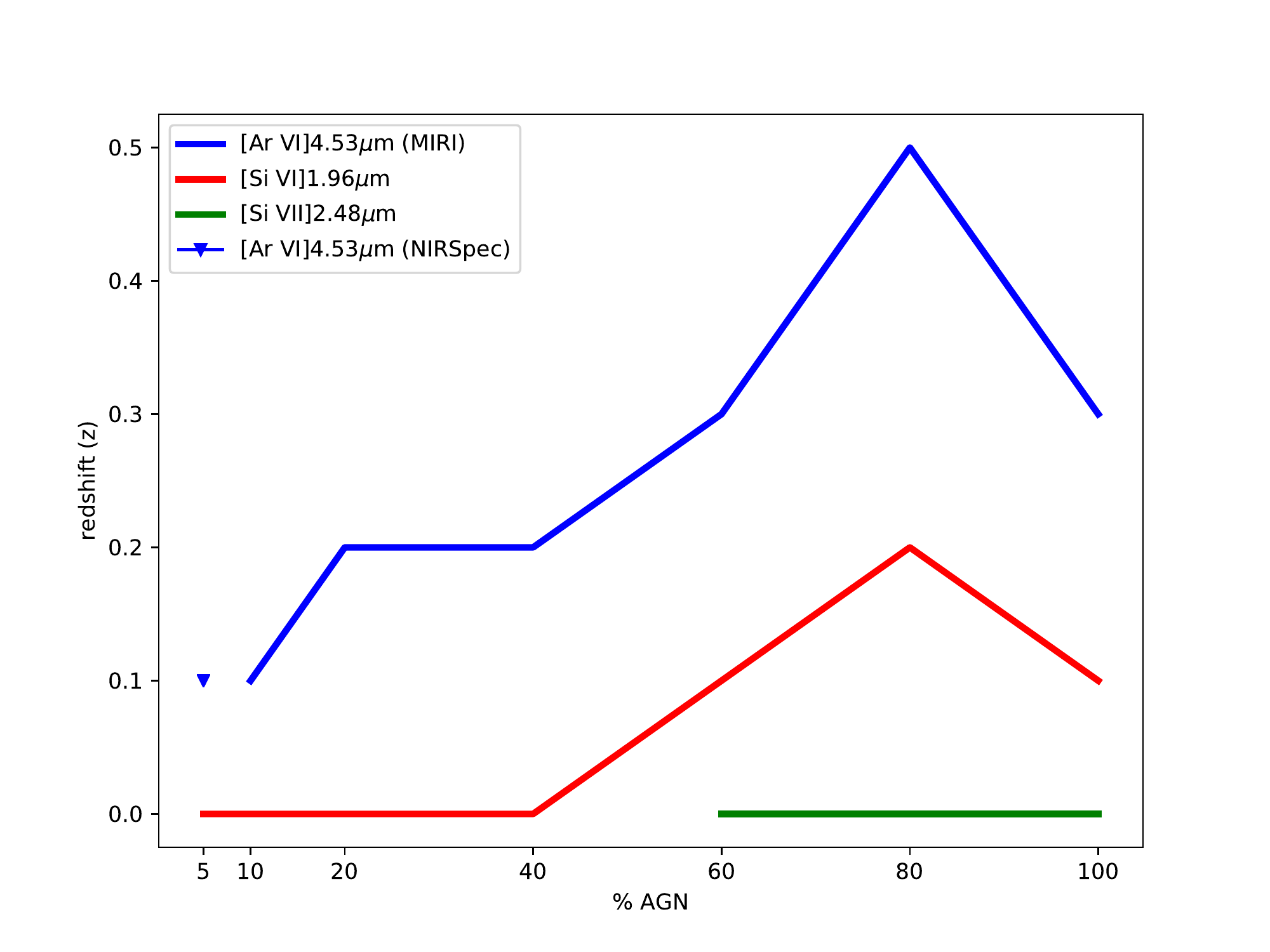}
\includegraphics[width=0.45\textwidth]{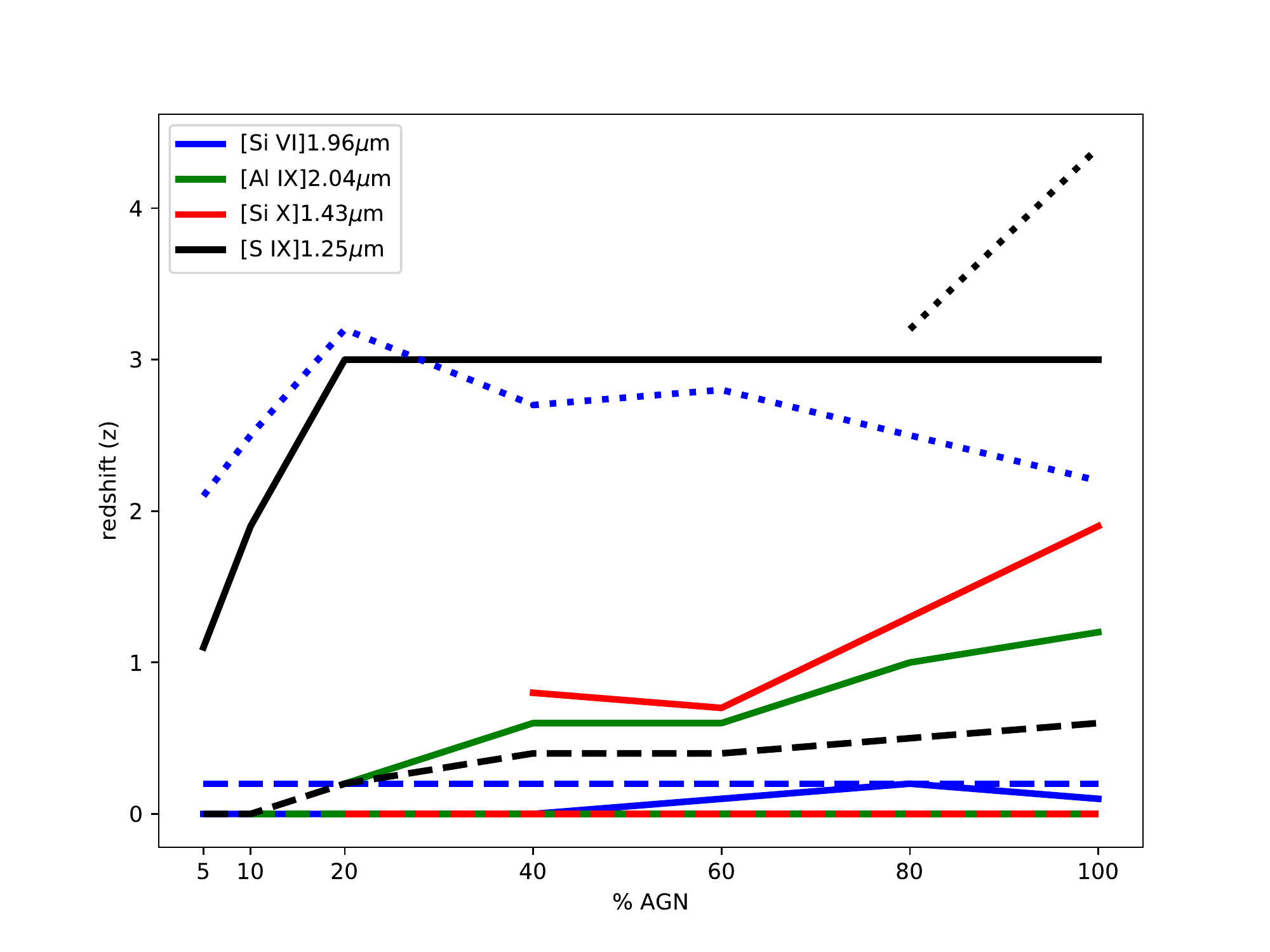}
\caption{Top Panel: Redshift at which key diagnostic lines with rest wavelengths in the NIRSpec wavelength range can be detected assuming predicted sensitivities as a function of \%AGN for our solar metallicity model with a standard gas density of $n_{\textrm{H}}=300~\rm{cm^{-3}}$, column density of $\log(N(H)/\textrm{cm}^{-2})=21$, and an ionization parameter of $\log U=-3.0$. From the ground, none of these lines can be detected by Gemini in this model, except for the [Ar~VI] line for $z< 0.1$ (blue triangle). For more dominant AGNs, the line can be detected by MIRI at redshifts out to 0.5.  Bottom Panel: Same as above but showing our $\log U=-1.0$ model, with all other parameters unchanged.  Solid lines corresponds to sensitivity limits of NIRSpec, dashed lines correspond to sensitivity limits of Gemini, and dotted lines correspond to MIRI sensitivities, which are used when the redshifted wavelength of the line moves into the MIRI band.}
\label{lum_nirspec}
\end{figure}

Figure \ref{lum_nirspec} shows the redshift at which the lines with rest wavelengths in the NIRSpec range can be detected as a function of \%AGN  for the same model as above but for a column density of $\log(N(H)/\textrm{cm}^{-2})=21$. Since the predicted line luminosities are so much lower in the NIRSpec range compared to those in the MIRI range, these diagnostic lines are only detectable for redshifts less than 0.1 for weakly accreting AGNs, and up to 0.5 for dominant AGNs, when the line redshifts into the MIRI range and is luminous enough to detect at wavelengths where the sensitivity is worse. For very weak AGN (\%AGN $<$ 10), the [Ar~VI] line is detectable only for redshifts $<$ 0.1 by NIRSpec (indicated by blue triangle in top panel of Figure \ref{lum_nirspec}) but is not bright enough to be detected by MIRI at higher redshifts. For dominant AGNs, the line is bright enough where it is detectable up to redshift 0.5 and shifted into the MIRI band.  Most notably, the top panel of Figure \ref{lum_nirspec} shows that the [Si~VI] line is detectable from the ground only for galaxies where the AGNs are dominant and only until redshifts of 0.2. It is therefore not surprising that this line is not always detected in ground-based observations of some local AGNs \citep{lamperti2017}. The detectability of the emission lines for our high ionization parameter model is shown in the bottom panel of Figure \ref{lum_nirspec}.  Here, solid lines corresponds to sensitivity limits of NIRSpec, dashed lines correspond to sensitivity limits of Gemini NIRI, and dotted lines correspond to MIRI sensitivities, which are used when the redshifted wavelength of the line moves into the MIRI band. As can be seen, at higher ionization parameters, the line luminosities are considerably larger, and for weak AGNs, the [Si~VI] line can be detected by MIRI out to redshift of 2, extending out to a redshift of 3 for more dominant AGN. For the most dominant AGNs, the [S~IX] line can be detected by MIRI out to a redshift of over 4. Note that [S~IX] is redshifted out of the NIRSPEC band at redshift 3, but does not get bright enough to be detected by MIRI until the AGN \% exceeds 80\%.  From Gemini (dashed lines), only the [Si~VI] and [S~IX] can be detected for weakly accreting AGNs for these high ionization parameter models out to a redshift of $< 0.2$.  At column densities of $\log(N(H)/\textrm{cm}^{-2})=23$, the near-infrared lines become significantly attenuated, and none of the lines can be detected from the ground at any ionization parameter. For low ionization parameters, the near-infrared CLs can only be detected in nearby galaxies with NIRSpec. At high ionization parameters, the [SiVI] line can be detected for dominant but heavily obscured AGNs out to redshift of only approximately 1. These figures highlight the crucial need for {\it JWST} in detecting these CLs beyond the very local universe. 

\begin{figure}

\includegraphics[width=0.45\textwidth]{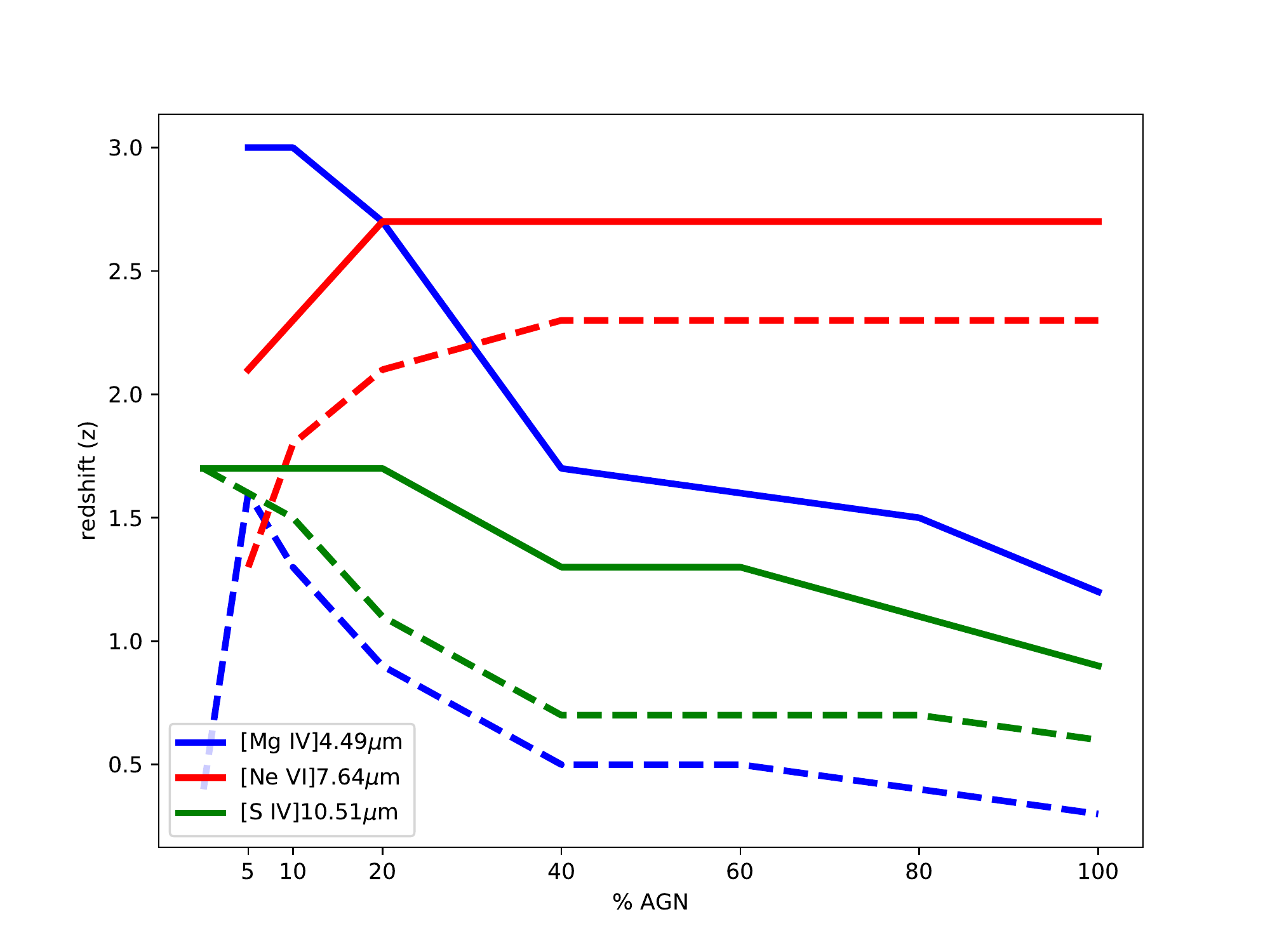}

\caption{Top Panel: Dependence of gas phase metallicity on redshift at which key diagnostic lines with rest wavelengths in the MIRI wavelength range can be detected assuming predicted sensitivities as a function of \%AGN for a standard gas density of $n_{\textrm{H}}=300~\rm{cm^{-3}}$, column density of $\log(N(H)/\textrm{cm}^{-2})=21$, and an ionization parameter of $\log U=-1.0$. Solid line corresponds to the Solar metallicity model while dashed lines correspond to our 0.1 Solar model.}
\label{lum_z}
\end{figure}

Note that the gas phase metallicity will have a significant impact on the detectability of the emission lines since lower metallicity environments will produce lower luminosity emission lines.  In Figure~\ref{lum_z}, we show the affect of metallicity on detectability of the representative lines our high U model. 

As can be seen, the reduced luminosity of the lines will have a considerable impact on the maximum redshift that can be detected. 

\section{Pre-selection Strategies for Identifying Elusive AGN Candidates}
As {\it JWST} is not a survey telescope, pre-selection strategies for selecting promising targets for follow-up study are invaluable. An ideal sample would consist of galaxies that are not optically-identified AGNs and are not luminous enough in the X-rays to be unambiguously identified as AGNs (e.g., $L_\mathrm{2-10~keV}>10^{42}$~erg~s$^{-1}$). While mid-infrared color selection can uncover optically-hidden and Compton-thick AGNs, widely used mid-infrared color cuts are generally too restrictive and unnecessarily exclude less luminous AGNs that are surrounded by significant star formation \citep{jarrett2011}, and would therefore only select the most dominant AGNs. In \cite{satyapal2018}, we constructed a theoretical mid-infrared color cut  for low redshift ($z < 0.2$) that will exclude even the most extreme starburst that we have modeled in this work. This cut is less restrictive than commonly employed, and can uncover twice the number of optically unidentified AGN candidates in the local Universe as shown in \citet{satyapal2018}. Here we provide a redshift-dependent selection criteria based on our models that can be applied out to a redshift of 1.0. In Cann et al.\ (2020, in preparation), we explore how these color cuts are affected when the black hole masses fall into the IMBH regime.

\begin{equation}
W1-W2 > \bigg\{\begin{array}{lr}
    -0.089 z + 0.6, & z \leq 0.45 \\
    2.496 z^2 - 3.874 z + 1.798, & 0.45 < z \leq 1
    \end{array} 
\end{equation}

\noindent The cut we provide here is more stringent and is designed to more reliably exclude pure starbursts, at the cost of some statistical completeness, in order to keep a pre-selected target list for non-survey facilities such as JWST to a reasonable size. In designing this cut, we restricted our starburst models to ionization parameters less than logU = -1.0 and column densities $\log(N(H)/\textrm{cm}^{-2})$ $<$ 22.5, in excess of which the optical continuum is attenuated by almost six orders of magnitude and the optical emission lines from the ionized gas are undetectable (see Satyapal et al. 2018). We note that our ionization parameter cut is even higher than the ionization parameters that are observed in the most extreme young low metallicity galaxies at high redshifts \citep{erb2010}. This allows us to make no cut on $W2-W3$, as the $W1-W2$ cut we provide here is sufficient to reliably exclude these extreme starbursts, and avoiding a cut on $W2-W3$ is advantageous for higher redshift sources where the S/N of the shallower $W3$ band is generally considerably less than $W1$ or $W2$.


\section{Conclusions}
We have conducted a theoretical investigation of the infrared emission line spectrum in the 1-30~$\micron$ range and mid-infrared spectral energy distribution (SED) produced by dust heated by an active galactic nucleus (AGN) and an extreme starburst. We used photoionization and stellar population synthesis models that extended past the ionization front in which both the line and emergent continuum is predicted from gas exposed to the ionizing radiation from a young starburst and an AGN. We note that the models we have presented here do not include the effects of shocks, which can affect both the standard optical spectroscopic diagnostics as well as the heating of the grains and the resulting transmitted mid-infrared SED. Our full suite of simulated SEDs are publicly available to download online  \footnote{\url{https://bgc.physics.gmu.edu/jwst-spectral-predictor/}} in a web-based form, where the users can select input parameters, visualize the spectra, immediately obtaine the 15 brightest emission lines, and download the predicted emission lines and continuum. We believe that this will be helpful to the community in preparation for {\it JWST} observations.

Our main results can be summarized as follows:


\begin{enumerate}
\item{We find that {\it JWST} is a powerful tool for uncovering and characterizing elusive AGNs when widely used optical, X-ray, mid-infrared color, and radio surveys fail. We show that low level accretion activity even in the nearby Universe cannot be detected by ground-based telescopes, underscoring the crucial need for {\it JWST} in the study of low level accretion activity.\\}

\item{We show that assuming typical parameters, our models predict that the [Mg IV] and [Ar VI]  lines near 4.5~$\mu$m are ideal lines to uncover elusive AGNs and can potentially be detected by MIRI out to a redshift of $\approx$ 1-3 for moderately luminous AGNs, depending on the ionization parameter and gas phase metallicity. For AGNs that dominate the bolometric luminosity, the [Si IX] line near 2.6~$\mu$m can be detected out to even higher redshift for a given ionization parameter and gas phase metallicity.\\}

\item{We show that \textit{JWST}, with the MIRI instrument in particular, is crucial to detect elusive AGNs that do not dominate the bolometric luminosity of the galaxy at all redshifts, and is crucial for detecting dominant AGNs beyond the local Universe (z $>$ 0.2).\\}

\item{We provide a redshift-dependent theoretical AGN color cut using the {\it WISE} bands that will exclude even the most extreme starburst. This color cut is less restrictive than other mid-infrared color-cuts employed in the literature which are designed to select luminous dominant AGNs, and can produce twice as many optically unidentified AGN candidates for follow-up studies.  \\}

\end{enumerate}

 We have demonstrated that infrared CL studies with {\it JWST} will offer an ideal opportunity to uncover an as yet undiscovered population of accreting black holes in the local Universe and beyond.

\section{Acknowledgements}

We thank the referee for their thoughtful comments. The simulations carried out in this work were run on ARGO, a research computing cluster provided by the Office of Research Computing at George Mason University, VA. (\url{ http://orc.gmu.edu})
This research made use of Astropy,\footnote{\url{http://www.astropy.org}} a community-developed core Python package for Astronomy \citep{astropy2013}, as well as \textsc{topcat} \citep{Taylor2005}.  J.M.C. gratefully acknowledges support from the National Science Foundation Graduate Research Fellowship Program.

This publication makes use of data products from the Wide-field Infrared Survey Explorer, which is a joint project of the University of California, Los Angeles, and the Jet Propulsion Laboratory/California Institute of Technology, and NEOWISE, which is a project of the Jet Propulsion Laboratory/California Institute of Technology. WISE and NEOWISE are funded by the National Aeronautics and Space Administration.

\facilities{Sloan, WISE}

\software{
Astropy \citep{astropy2013}, 
\textsc{topcat} \citep{Taylor2005}
}




\end{document}